\documentclass[12pt]{article}
\usepackage{amssymb,latexsym,amsmath} 
\usepackage{authblk}
\usepackage[utf8]{inputenc}
\usepackage{mathtools}
\usepackage{epstopdf}
\usepackage{dsfont}
\usepackage{graphics}
\usepackage{color}
\usepackage{subfig}
\usepackage{graphicx}
\usepackage{mathrsfs}
\usepackage{placeins}
\usepackage{empheq}
\usepackage{siunitx,array,multirow}
\usepackage[colorlinks = true,
linkcolor = blue,
urlcolor  = blue,
citecolor = blue,
anchorcolor = blue]{hyperref}
\usepackage[numbers,square,sort&compress]{natbib}

\usepackage{pifont}
\usepackage{multirow}
\usepackage[width=\textwidth]{caption}

\begin{document}
%-----------------------------------------------------------------
\title{\bf{Generalized complexity and dynamical response in holographic Vaidya spacetimes}}

\author[1]{Mojtaba Shahbazi \thanks{mojtaba.shahbazi@modares.ac.ir}}
\author[2]{Monireh Emami \thanks{monireh.emami@ipm.ir}}

\affil[1]{Department of Physics, Faculty of Basic Sciences, Ayatollah Boroujerdi University, Boroujerd, Iran}
\affil[2]{School of Particles and Accelerators, Institute for Research in Fundumental Sciences (IPM), Tehran, Iran}
\maketitle
%----------------------------------------------------------------
\begin{abstract} 
We investigate "Complexity = Anything" for smooth Vaidya geometries, using the Weyl squared functional as the most common candidate for our study. Our numerical analysis of candidates in 4- and 5-dimensional Reissner–Nordström (RN) and 5-dimensional Gauss-Bonnet (GB) shows evolution in both $r$ and $v$ ("doubled complexity"), unlike static solutions that depend only on $r$. We then study the difference between the complexity of static and dynamical solutions in the asymptotically AdS regime using a Fefferman-Graham expansion. By expanding the bulk metric, extremal embedding, induced metric, normal vector, and extrinsic curvature simultaneously, we show that the first four FG coefficients cancel between the two geometries. In contrast, the first nonvanishing contribution occurs at the fifth coefficient and is controlled by the boundary stress tensor and its derivatives. During the Vaidya quench, the derivative contribution encodes the time-dependent response of the boundary state. In linear response, this response is governed by the retarded stress-tensor correlator and, through generalized Kramers-Kronig relations, can be represented in terms of its spectral density. We thus identify a connection between generalized holographic complexity and the dynamical stress-tensor response, which in the linear-response regime can be represented in terms of the corresponding stress-tensor spectral density.
\end{abstract}

%--------------------------------------------------------------------
\section{Introduction}

Holography gives a geometric description of quantum computational complexity. The complexity of a boundary state may be characterized, in a quantum circuit formulation, by the minimum number of elementary gates required to prepare that state from a reference state. In holography, several geometric prescriptions have been proposed for this quantity, including complexity=volume (CV) \cite{cv}, complexity=action (CA) \cite{ca}, and the volume of the Wheeler-DeWitt patch \cite{cv2}. More recently, the complexity=anything (CAny) proposal has provided a generalized framework in which complexity is associated with a broad class of diffeomorphism-invariant bulk observables \cite{general,cany2}. In particular, codimension-one functionals constructed from curvature invariants allow geometric structures beyond the standard volume functional to contribute to the complexity.

In several works, $\mathcal{C}_{Any}$ has been studied for static $(d+1)$-dimensional metric solutions for Einstein and beyond gravity equations, e.g., \cite{omidi,rn,gbcom,pht,mp,love}. Moreover, the holographic complexity in Vaidya geometry with shock-wave energy injection has been studied previously within standard complexity proposals, for example in \cite{vaidya1, vaidya2, jiang, shsa, ali} and CAny \cite{vaidya3}, with massless and massive black hole solutions before insertion. Here, we investigate this observable for the Vaidya geometry with a smooth time change from initial zero mass and other time-dependent parameters, adding a new dimension: the complexity exhibits dynamics. More precisely, we study the $ r $-constant slices and find that this represents the complexity quantity similarly to the complexity of $ v $-constant (the static cases). We refer to it as ``doubled complexity". We use the term doubled description to denote the two-parameter dependence of the extremal hypersurface on $(r,v)$, rather than proposing a new independent holographic complexity. We study the CV proposal and CAny to see how the generalization of the complexity functional modifies this dynamical behavior.

The presence of curvature dependent structures in the complexity functional raises a natural question: what information about the boundary quantum state is encoded in the corresponding dynamical complexity? This question is especially interesting for Vaidya geometry, a time-dependent process in which energy is injected into the boundary system.

In this work, we focus on a specific and common candidate generalization function: the square of the Weyl tensor. We study the corresponding generalized complexity in charged Vaidya solutions within Einstein theory and beyond. Because we discuss their effects on the boundary limit, the higher-dimensional solutions guarantee nonzero terms in the curvature scalar expansions in that limit. Our numerical studies concern 4- and 5-dimensional cases. Because Lovelock theory is significant in higher dimensions as a generalization of Einstein's, we present the Gauss-Bonnet (GB) term as the first term of this generalization. This example, in 5 dimensions, the minimum dimension in which this term contributes dynamically, can also be a follow-up of one of our previous papers \cite{love} in the context of CAny for charged static Lovelock spacetimes.

We then consider a more general question that does not depend on the detailed form of the bulk matter sector. We compare the generalized complexity of the dynamical Vaidya geometry with that of the corresponding static geometry and study their difference using an FG expansion. The expansion of the complexity functional requires the simultaneous expansion of the bulk metric, the extremal embedding, the induced metric, the normal vector, and the extrinsic curvature. Following the FG treatment of extremal codimension-one surfaces \cite{faraj,comment}, we show that the first several coefficients (up to the fourth order) in the FG expansion of the two complexities coincide. This cancellation is not a consequence merely of the equality of the boundary metric, i.e., it follows from the simultaneous expansion of the bulk geometry and the embedding data. The first nonvanishing term contribution contains the coefficient $g^{(4)}_{\mu\nu}$, which is related holographically to the expectation value of the boundary stress tensor \cite{renor}.

Consequently, the leading nonvanishing term contribution to the complexity difference contains terms of the form $T_d-T_s$ and $\nabla T_d$, where $T_d$ is the energy-momentum tensor of the dynamical case, and $T_s$ of the static case. For the Vaidya collapse, the difference between the static and dynamical stress tensors is suppressed in the thin-shell region, leaving the derivative of the time-dependent stress tensor as the relevant contribution. Thus, this complexity difference provides a geometric probe of the dynamical response generated by energy injection. The relation to response theory follows from the boundary stress-tensor response. By generalized Kramers-Kronig (KK) dispersion formulas \cite{kk}, the response can be represented as a functional of the spectral density. For the time-dependent quench considered here, we use the generalized response theory form of the dispersion relations rather than assuming time translation invariance of the perturbed state. Consequently, the time derivative of the boundary stress tensor entering the leading contribution to the complexity difference can be related to the stress tensor spectral density, where the precise frequency kernel is determined by the corresponding response relation. Thus, the complexity difference yields a geometric quantity sensitive to the boundary theory spectral response.

The paper is organized as follows. In the section \ref{vaidya}, we study the generalized complexity in the Vaidya geometry and investigate the resulting complexity growth rate from $ r $-constant slices. In the section \ref{psec}, we perform the FG analysis of the difference between the static and dynamical complexities and relate them to the boundary stress tensor and dynamical response. Our conclusions are presented in the final section.
%-----------------------------------------
\section{Complexity in AdS-Vaidya spacetime}\label{vaidya}
In the AdS-Vaidya spacetime, complexity is not limited to a specific geometric quantity but can be described by a more general function that includes a class of new diffeomorphism invariant observables \cite{general, cany2}. Here we concentrate on the codimension-one case for general $ (d+1)-$dimensional bulk regions \cite{general,cany2}
\begin{align}\label{obser}
	\mathcal{O}_{\mathcal{F}_1,\Sigma_{\mathcal{F}_2}}(\Sigma_{CFT})=\frac{1}{G_N \ell}\int_{\Sigma_{\mathcal{F}_2}}\mathrm{d}^{d}\sigma \sqrt{h} \mathcal{F}_1(g_{\mu\nu};X^{\mu})
\end{align}
where, $\mathcal{F}_1$ can be set to one, which leads to the volume of the hypersurface as in CV. It can also be a general scalar function of the metric $g_{\mu\nu}$ and of an embedding $X^{\mu}(\sigma^a)$ of the hypersurfaces. Furthermore, $\Sigma_{\mathcal{F}_2}$ is a codimension-one hypersurface in the bulk spacetime with boundary time slice $\partial\Sigma_{\mathcal{F}_2} =\Sigma_{CFT}$. Extremality of the hypersurface leads to \cite{general,cany2}
\begin{align}\label{var}
	\delta_{X}\Big(\int_{\Sigma}\mathrm{d}^{d}\sigma \sqrt{h} \mathcal{F}_2(g_{\mu\nu};X^{\mu})\Big)=0. 
\end{align}
For simplicity, we follow the case $\mathcal{F}_1 =\mathcal{F}_2$, so the observable \eqref{obser}, known as $\mathcal{C}_{Any}$, obeying the above condition is expressed by \cite{general,cany2}
\begin{align}\label{maxv}
	\mathcal{C}_{Any}(\tau)= \max_{\partial\Sigma(\tau)=\Sigma_{\tau}}\frac{V_0}{G_N \ell}\left[\int_{\Sigma}\mathrm{d}^{d}\sigma \sqrt{h} \mathcal{F}_1(g_{\mu\nu};X^{\mu})\right]
\end{align}
where $h$ is the determinant of the  induced metric on the given hypersurface.
A usual choice of generalization is
\begin{align}\label{F1F2}
	\mathcal{F}_1=\mathcal{F}_2=1+\gamma \ell^4 C^2,
\end{align}
where $C^2$ is the Weyl tensor squared and for $\gamma=0$ the CV result is recovered.

We take the metric in Eddington-Finkelstein coordinates
\begin{align}\label{metricv}
	ds^2=-f(r,v)\mathrm{d}t^2+\frac{1}{f(r,v)}\mathrm {d}r^2+\big(\frac{ r}{\ell}\big)^2 \mathrm {d} \Omega _{k,d-1}^2, 
\end{align}

The generalized volume complexity as a codimension-one observable can be obtained from \eqref{maxv} and \eqref{F1F2} as \cite{general,cany2}
\begin{align}\label{Any}
	\mathcal{C}_{Any}(\tau)= \frac{V_0}{G_N \ell}\int_{\Sigma}\mathrm{d}\sigma  \, \big(\frac{ r}{\ell}\big)^{d-1}\sqrt{-f(r,v) \dot{v}^{2}+2\dot{v} \dot{r}} \,a(r,v)
\end{align}
Here, the dots indicate derivatives with respect to $\sigma$, and $V_0$ represents the volume of the spatial directions of the $(d-1)$-dimensional submanifold.

Considering $\mathcal{C}_{Any}$ as an action, where extremizing it is equivalent to solving its equations of motion, leads to a conserved  momentum conjugate to the coordinate $v$, since spacetime is stationary: 
\begin{align}\label{pv}
	P_v=-\frac{\partial \mathcal{L}}{\partial \dot{v}}=\frac{a(r,v)(r/\ell)^{d-1}(\dot{r}-f(r,v) \dot{v})}{\sqrt{-f(r,v) \dot{v}^{2}+2 \dot{v} \dot{r}}}=\dot{r}-f(r,v) \dot{v}.	
\end{align}
Since \eqref{Any} is diffeomorphism invariant and does not change under reparametrization, one can fix the parameter $\sigma$ by choosing 
\begin{align}
	\sqrt{-f(r,v) \dot{v}^{2}+2 \dot{v} \dot{r}}=a(r,v) \big(\frac{ r}{\ell}\big)^{d-1}.
\end{align}
Then it is straightforward to derive the extremality conditions
\begin{align}\label{rdot}
	\dot{r}&=\pm \sqrt{P^2_v +f(r,v)a(r,v)^2\big(\frac{r}{\ell}\big)^{2(d-1)}},\\
	\label{tau}
	\dot{t}&=\dot{v}-\frac{\dot{r}}{f(r,v)}=\frac{-P_v\dot{r}}{f(r,v) \sqrt{P^2_v + f(r,v)a(r,v)^2(r/\ell)^{2(d-1)}}}.
\end{align}
Then, \eqref{rdot} takes the form of the equation of motion for a classical particle:	
\begin{align}\label{eom}
	\dot{r}^{2}+U(r,v)=P^2_v,
\end{align}
where $U(r)$ is the effective potential
\begin{align}
	U(r,v)=-f(r,v)\,a(r,v)^2\big(\frac{r}{\ell}\big)^{2(d-1)}.
\end{align}
For the symmetric trajectory, the conserved momentum is a function of the turning point $r_{min}$
\begin{align}
	P^2_v=	U(r_{min},v)=-f(r_{min},v)\,a(r_{min},v)^2\big(\frac{r_{min}}{\ell}\big)^{2(d-1)}.
\end{align}	

The boundary time derivative of $\mathcal{C}_{Any}$ leads to 
\begin{align}
	\frac{\mathrm{d} \mathcal{C}_{Any}}{\mathrm{~d} \tau}=\frac{1}{2} \frac{\mathrm{d} \mathcal{C}_{Any}}{\mathrm{~d} \tau_{\mathrm{R}}}&=\frac{V_0}{G_N \ell} P_v\nonumber\\
	&=\frac{V_0}{G_N\ell} \sqrt{-f(r_{min},v)}\,a(r_{min},v)\big(\frac{r_{min}}{\ell}\big)^{d-1}.
\end{align}
So, the growth rate of generalized complexity can be studied by the behavior of the conserved momentum over full time. In particular, the late-time behavior of the growth rate is determined by
\begin{align}\label{tauinf}
	\lim _{\tau \rightarrow \infty} \frac{\mathrm{d} \mathcal{C}_{Any}}{\mathrm{~d} \tau}=\frac{V_0}{G_N\ell} \sqrt{-f\left(\tilde{r}_{\min },v\right)} a(\tilde{r}_{\min },v)\big(\frac{\tilde{r}_{min}}{\ell}\big)^{d-1},
\end{align}
where $\tilde{r}_{\min }$ is the local maximum of the effective potential, and because at this radius time goes to infinity, it is sometimes known as $r_f\equiv \tilde{r}_{\min }$.

According to the analyze of \cite{vaidya1}, due to the dependence of $ P $ on $ v $, the second order derivatives of \eqref{rdot} and \eqref{tau} could be taken, leading to the full equations of motion
\begin{align}\label{rdr}
	\ddot{r}& = \frac{1}{2} \partial_v f \,\dot{v}^2 + \frac{1}{2} \partial_r \left( a^2 r^{2(d-1)} f \right),\nonumber\\\nonumber\\ 
	\ddot{v}& = \frac{1}{2} \partial_r \left( a^2 r^{2(d-1)} \right) - \frac{1}{2} \partial_r  f\, \dot{v}^2.
\end{align}
Derivation of \eqref{pv} (for the dynamical case) and comparing to \eqref{rdr} leads to
\begin{equation}\label{eq:Pdot}
	\dot P=-\frac{1}{2}\,\partial_v f\,\dot v^{2}
\end{equation}
which means that $P$ fails to be conserved only through the explicit $v$-dependence of the metric function, as expected. Setting $\partial_v f=0$ recovers $P=const.$ and the static analysis.

One can recover the first-order $ \sigma $-independent equations \cite{vaidya1} 
\begin{align}\label{dvdr}
	\frac{dv}{dr}&=\frac{1}{f}\left(1-\frac{P}{R}\right),
\end{align}
\begin{align}\label{dPdr}
	\frac{dP}{dr}&=-\frac{\partial_v f}{2\,f^{2}}\,\frac{(R-P)^{2}}{R},
\end{align}
where
\begin{equation}\label{Rdef}
	R(r,v,P)\;\equiv\;\dot r\;=\;\varsigma\sqrt{P^{2}-U(r,v)},\qquad \varsigma=\pm1.
\end{equation}

Integrating \eqref{tau} also for this case leads to a boundary time similar to the static case, but with functions of radius and time
\begin{align}\label{taup}
	\tau = -2 \int^{\infty}_{r_{min}} dr \frac{P(r_{min},v)}{f(r,v)\,\sqrt{P^{2}(r_{min},v)-U(r,v)}}.
\end{align}
This relation gives the time behavior of the momentum and in turn the complexity rate.

For Vaidya-type spacetimes with smooth functions of time, the above relation shows that the boundary time $ \tau $ depends on two variables, $ r_{min} $ and $ v $. Calculating this quantity with respect to both of them at once is subtle. Instead, we choose to do it by fixing one of them and varying the other one. In other words, we can calculate the boundary time upon the momentum on $v$-constant and $r$-constant slices. The first one is nothing but a static case, which has been studied several times. However, the latter, for a smooth range of time $v$ and fixed minimal radius, is a new insight of consideration.
 
Since we continue our discussion in the next section with the boundary limit, the higher-dimensional solutions guarantee nonzero terms in the curvature scalar expansions in that limit. We study numerically the 4- and 5-dimensional cases. For this purpose, we focus on charged Vaidya black branes in Einstein theory, Reissner–Nordström (RN), and the Gauss-Bonnet (GB), the first term of Lovelock theory as a significant generalization of Einstein theory in higher dimensions.

\begin{figure}[!htbp]
	\centering
	\centering
	\subfloat[]{\includegraphics[width=8cm]{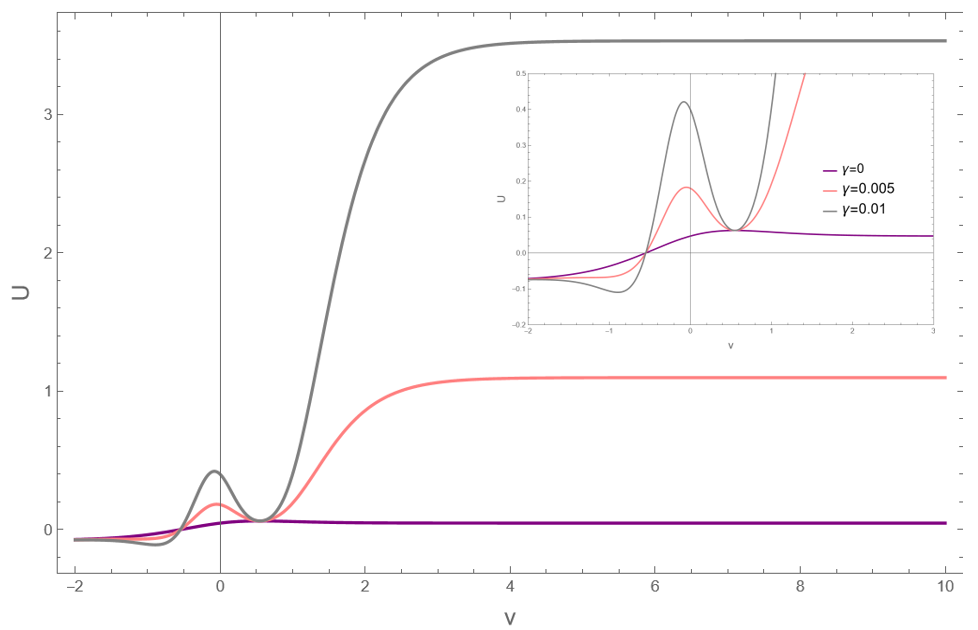} \label{fig:1a}}%
	\subfloat[]{\includegraphics[width=8cm]{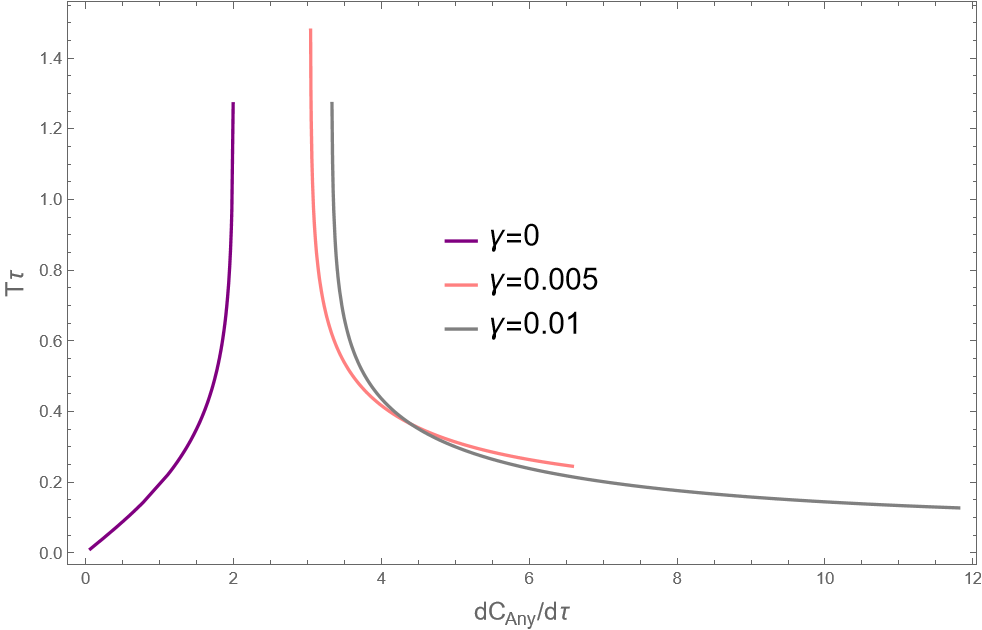} \label{fig:1b}}%
	\qquad
	\subfloat[]{\includegraphics[width=5.33cm]{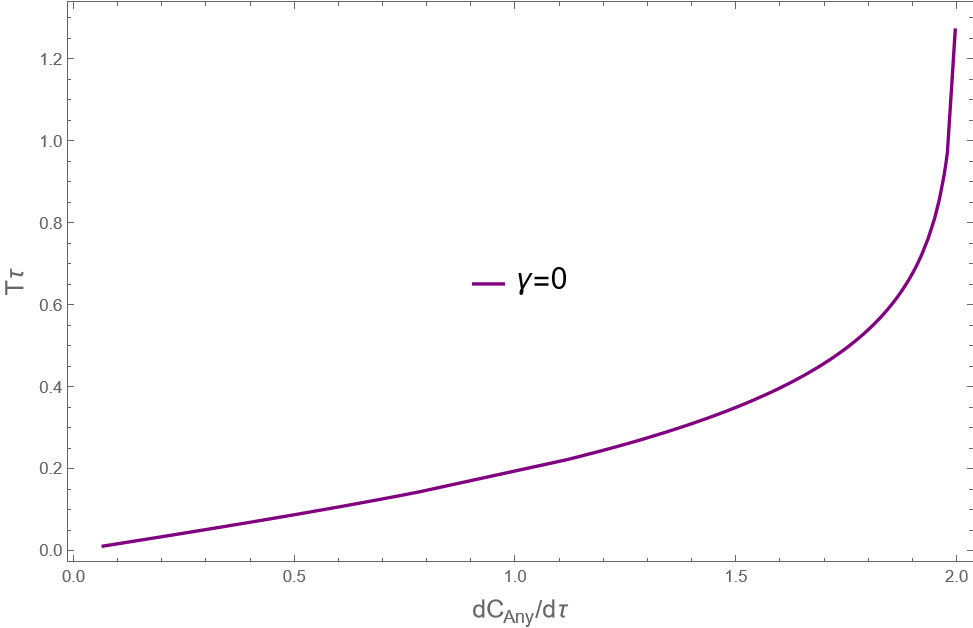} \label{fig:1c}}%
	\subfloat[]{\includegraphics[width=5.33cm]{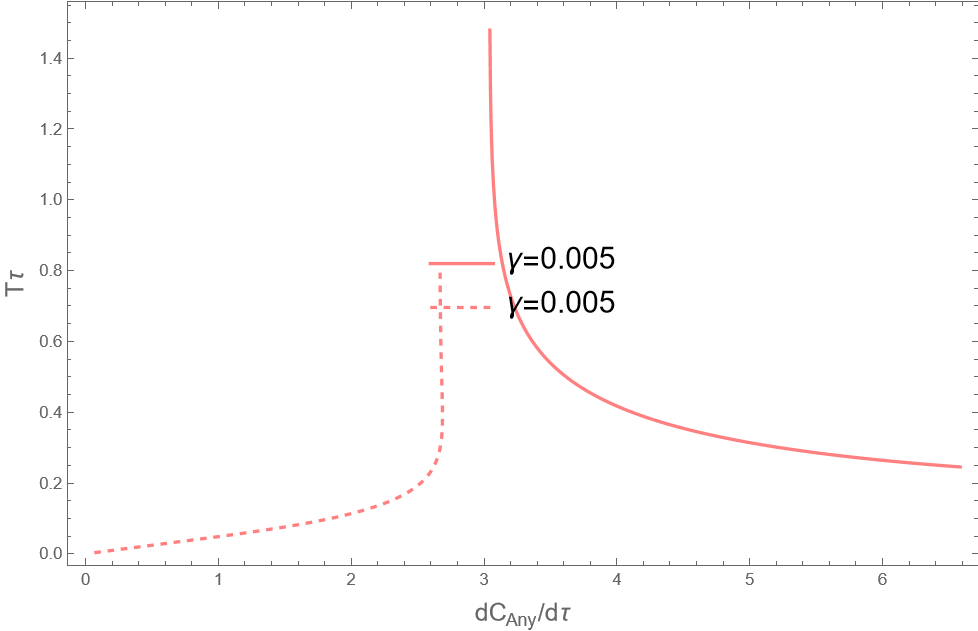} \label{fig:1d}}%
	\subfloat[]{\includegraphics[width=5.33cm]{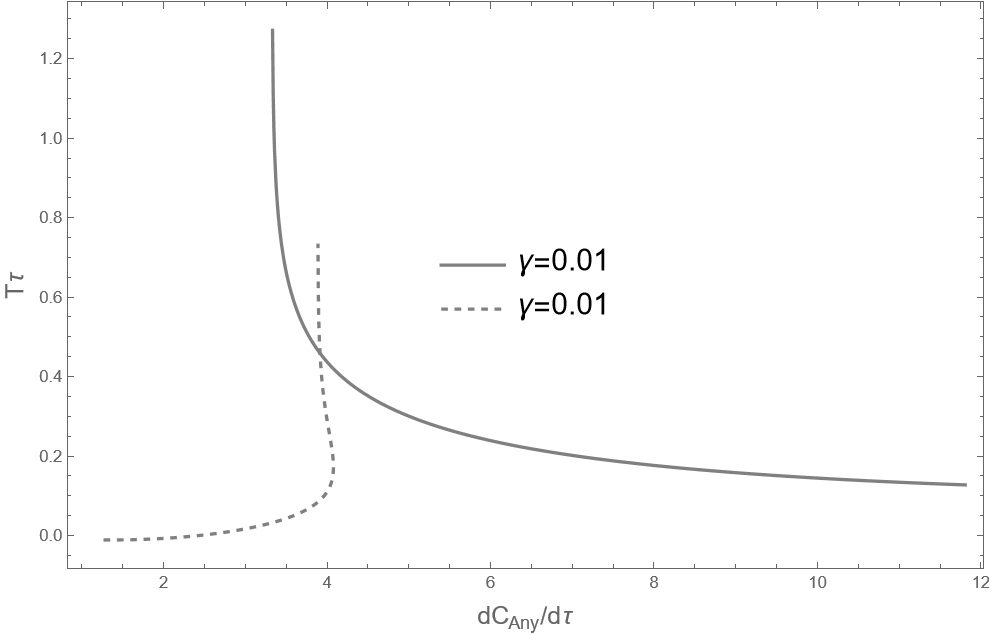} \label{fig:1e}}%
	\qquad
	\subfloat[]{\includegraphics[width=5.33cm]{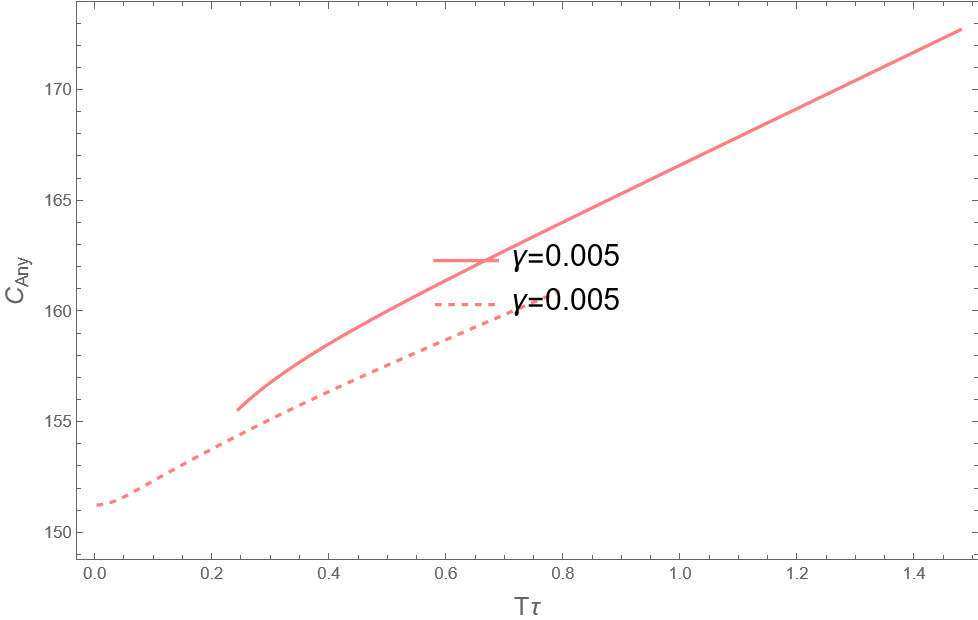} \label{fig:1f}}%
	\subfloat[]{\includegraphics[width=5.33cm]{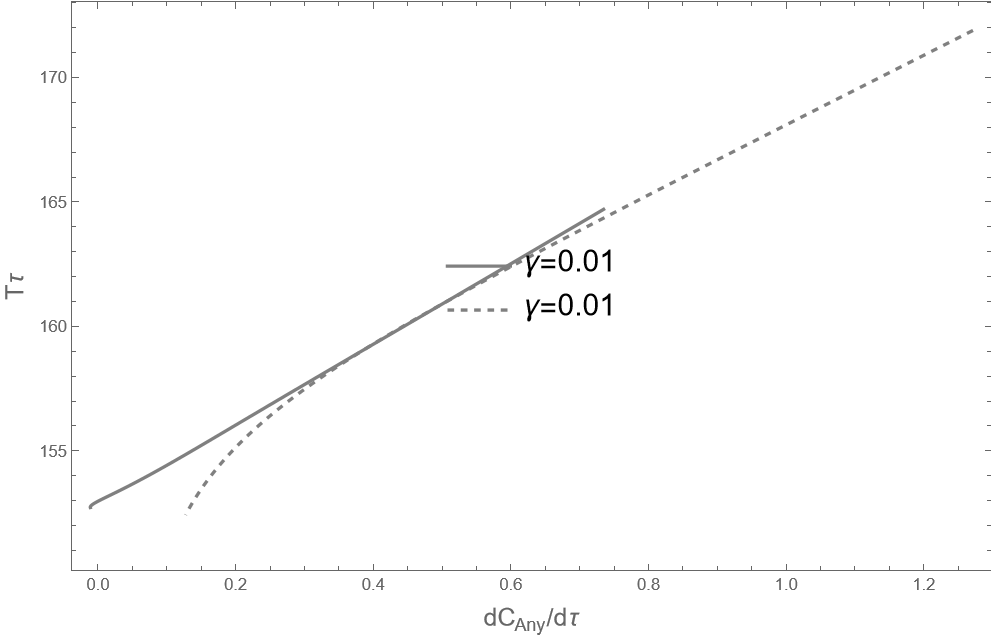} \label{fig:1g}}%
	\caption{(a) Effective potentials $U$ vs. radius $r$. (b--e) Boundary time multiplied by temperature vs. complexity time rate. (f, g) Complexity vs. boundary time multiplied by temperature for the 4-dimensional RN black brane with $r_{min}/r_+=0.5$, $\ell/r_+ = 1$, and $\tilde{Q}/r_+ = 1$. The temperature $T$ and the outer horizon radius $r_+$ are evaluated in the limit $v\to\infty$. The dashed curves in the second and third rows mark the portions excluded by the non-maximality of the complexity.}\label{fig:1}%
\end{figure}
\begin{figure}[!h]
	\centering
	\subfloat[]{\includegraphics[width=8cm]{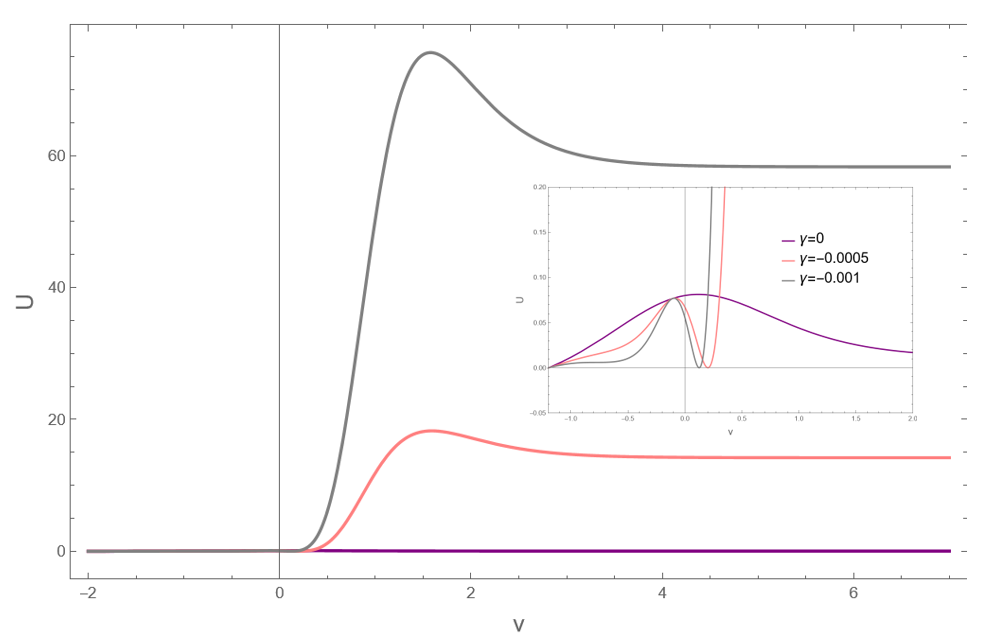} \label{fig:2a}}%
	\subfloat[]{\includegraphics[width=8cm]{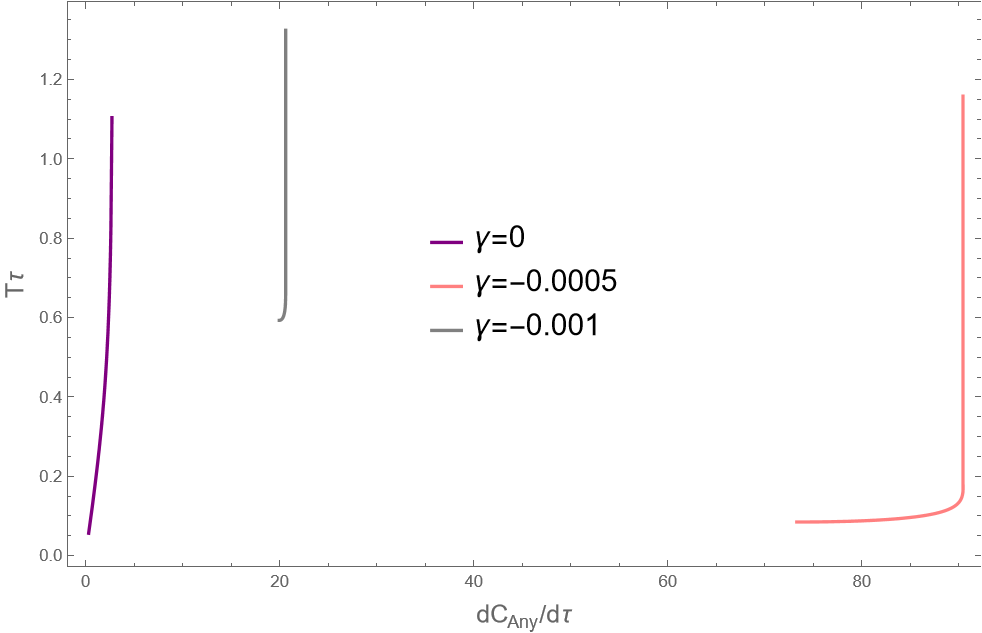} \label{fig:2b}}%
	\qquad
	\subfloat[]{\includegraphics[width=5.33cm]{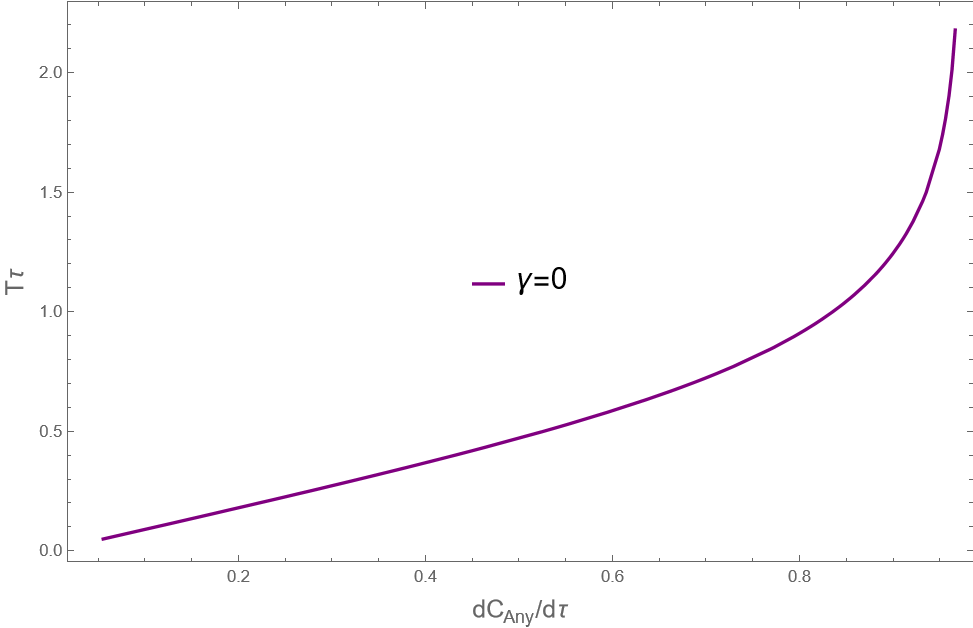} \label{fig:2c}}%
	\subfloat[]{\includegraphics[width=5.33cm]{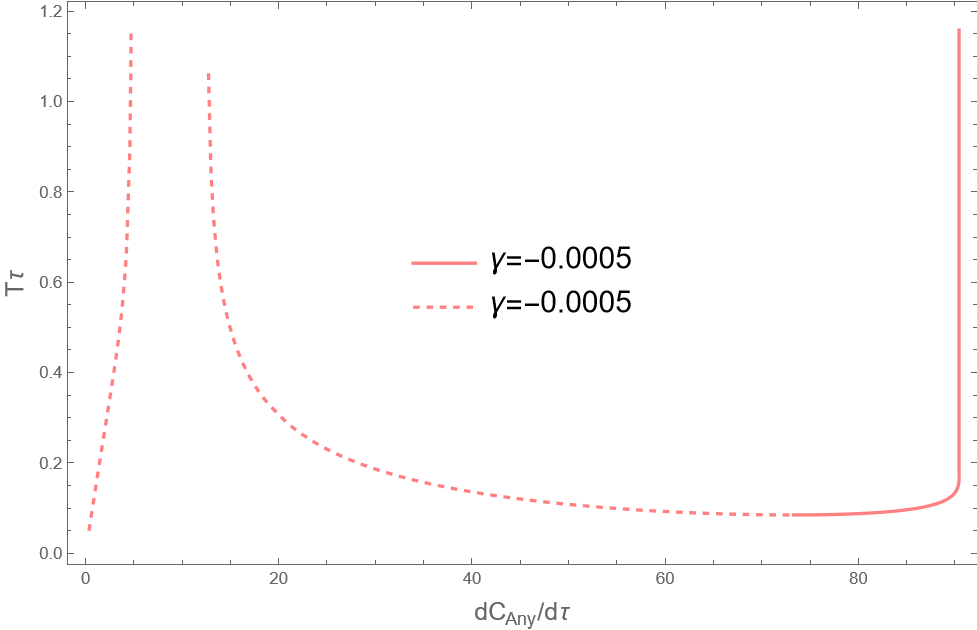} \label{fig:2d}}%
	\subfloat[]{\includegraphics[width=5.33cm]{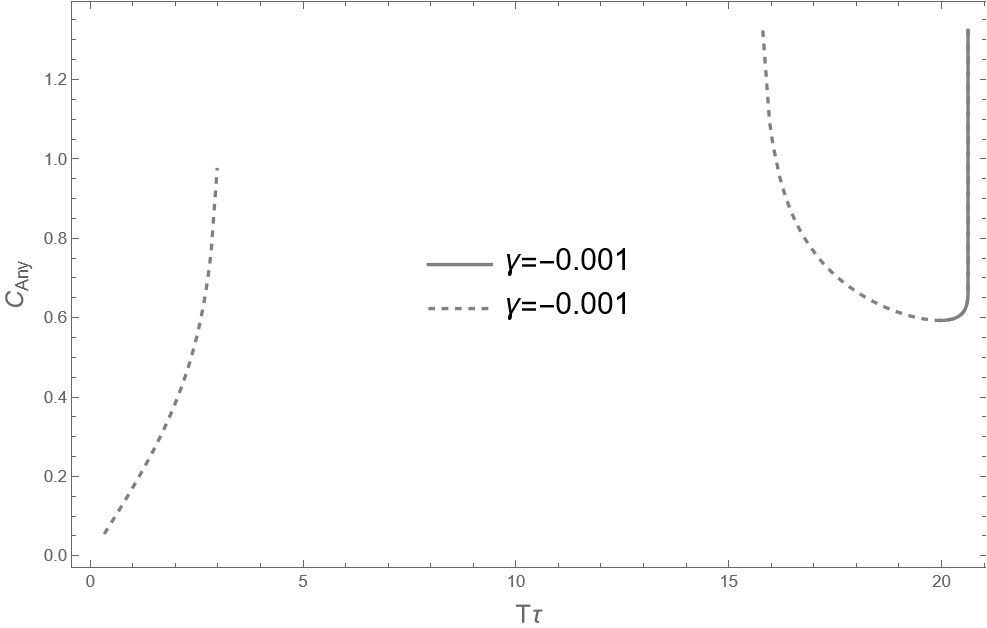} \label{fig:2e}}%
	\qquad
	\subfloat[]{\includegraphics[width=5.33cm]{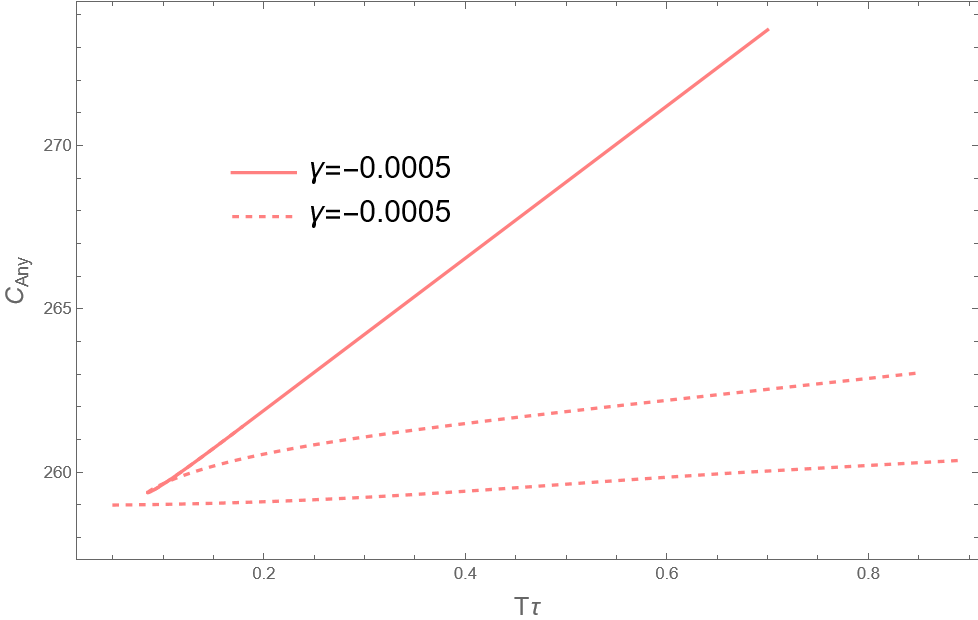} \label{fig:2f}}%
	\subfloat[]{\includegraphics[width=5.33cm]{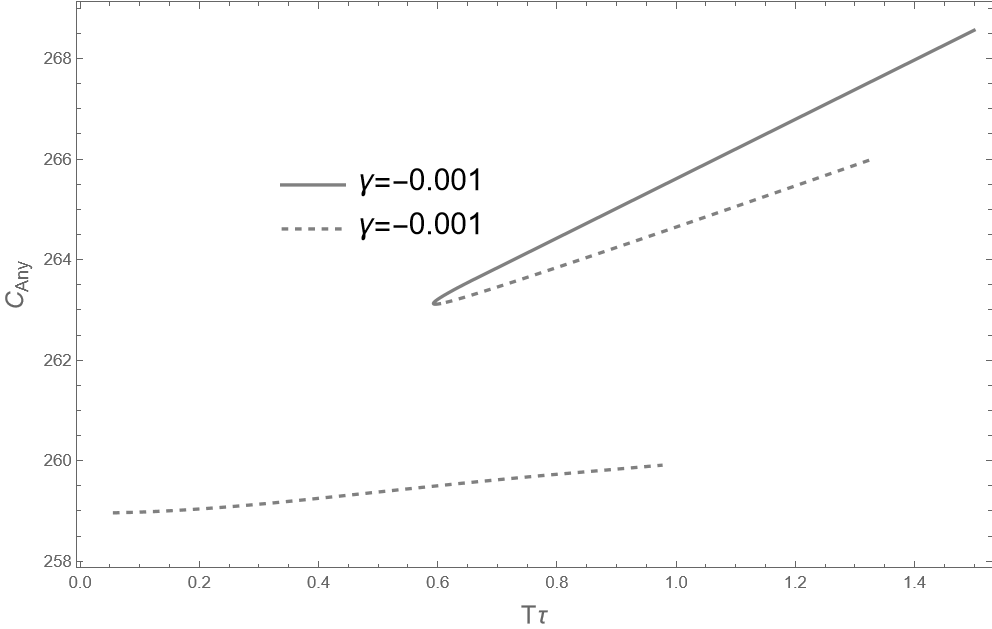} \label{fig:2g}}%
	\caption{(a) Effective potentials $U$ vs. radius $r$. (b--e) Boundary time multiplied by temperature vs. complexity time rate. (f, g) Complexity vs. boundary time multiplied by temperature for the 5-dimensional RN black brane with $r_{min}/r_+=0.5$, $\ell/r_+ = 0.5$, and $\tilde{Q}/r_+^2 = 1.2$. The temperature $T$ and the outer horizon radius $r_+$ are evaluated in the limit $v\to\infty$. The dashed curves in the second and third rows mark the portions excluded by the non-maximality of the complexity.}\label{fig:2}%
\end{figure}
\begin{figure}[!h]
	\centering
	\subfloat[]{\includegraphics[width=8cm]{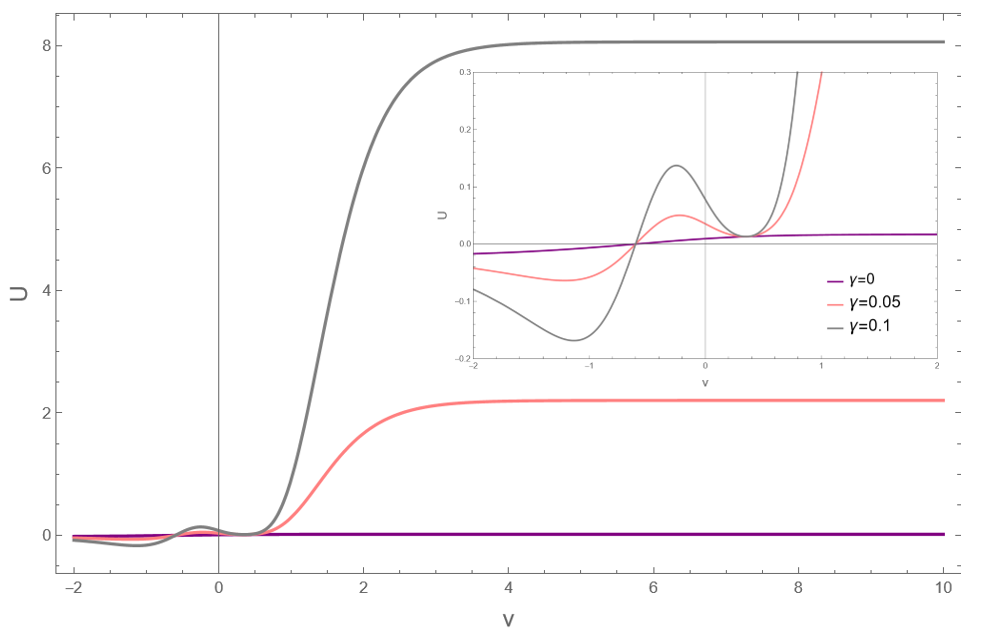} \label{fig:3a}}%
	\subfloat[]{\includegraphics[width=8cm]{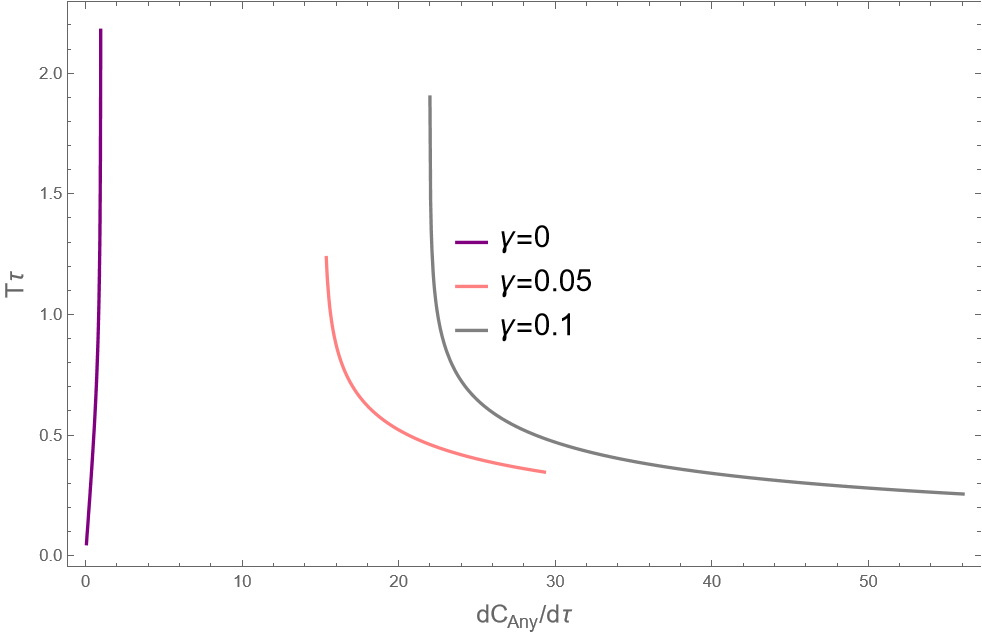} \label{fig:3b}}%
	\qquad
	\subfloat[]{\includegraphics[width=5.33cm]{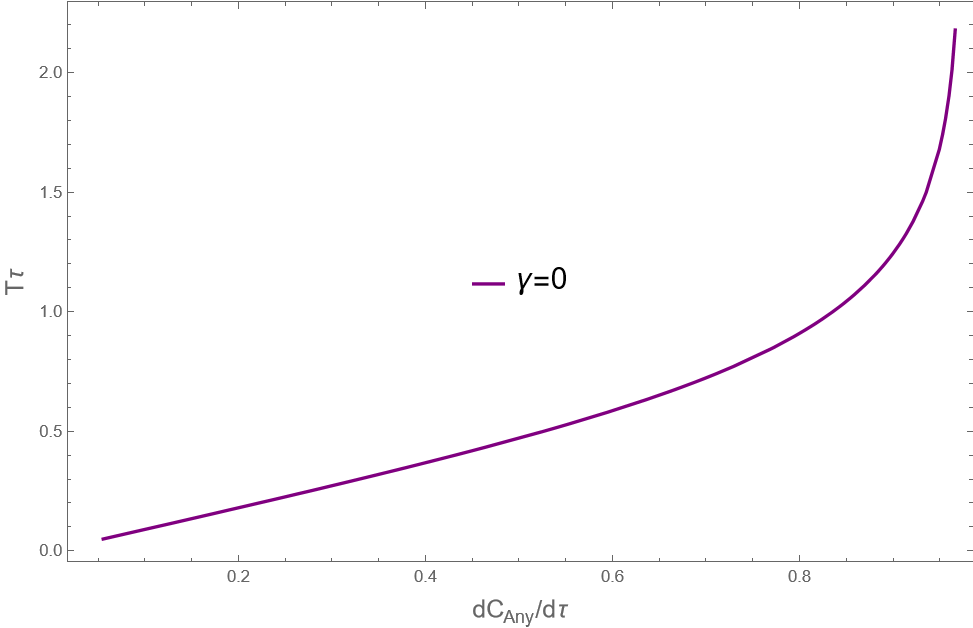} \label{fig:3c}}%
	\subfloat[]{\includegraphics[width=5.33cm]{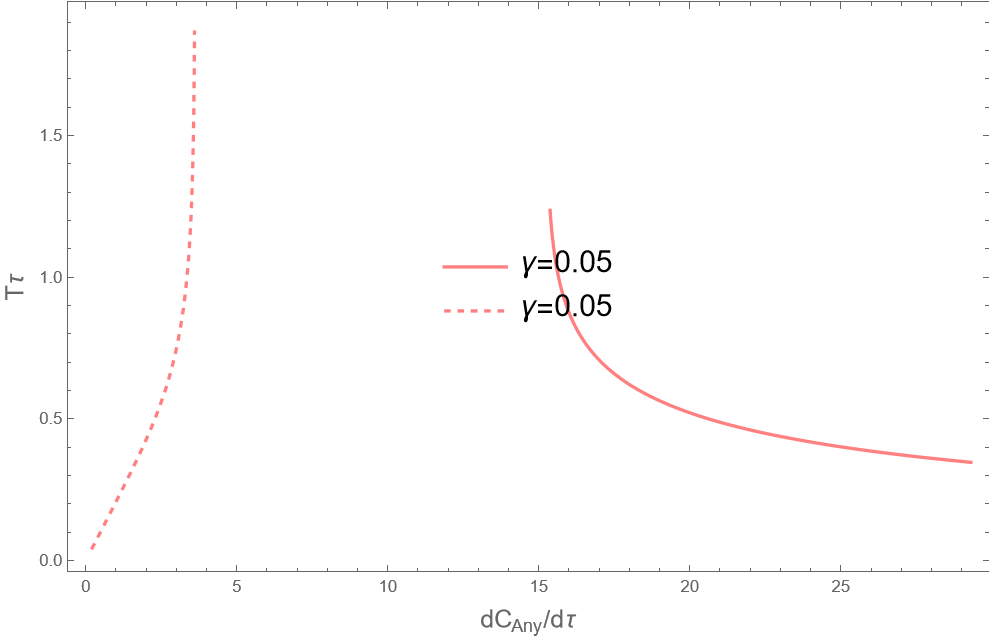} \label{fig:3d}}%
	\subfloat[]{\includegraphics[width=5.33cm]{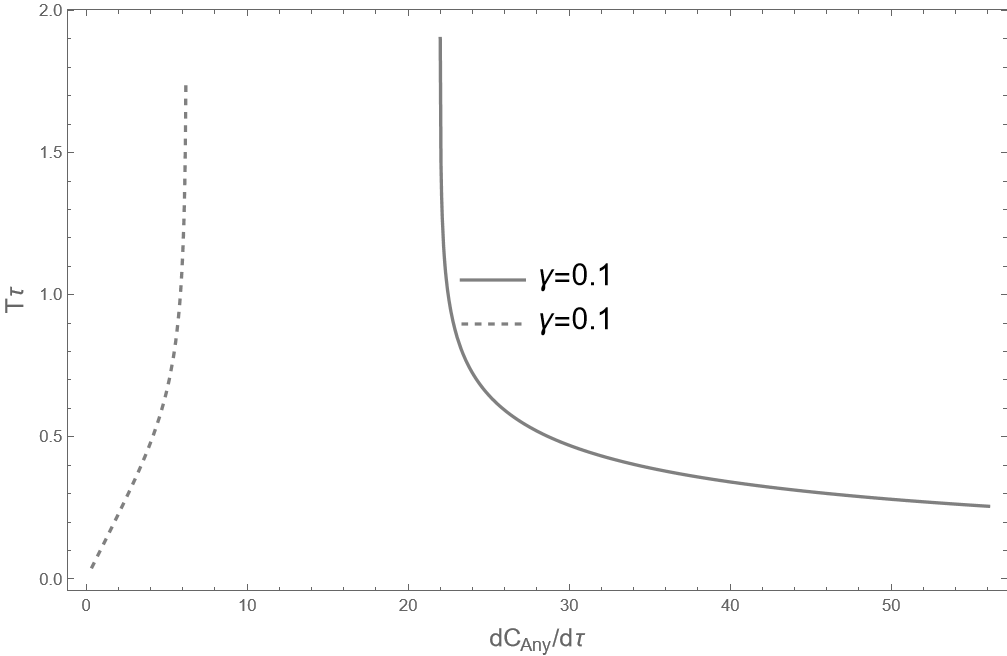} \label{fig:3e}}%
	\qquad
	\subfloat[]{\includegraphics[width=5.33cm]{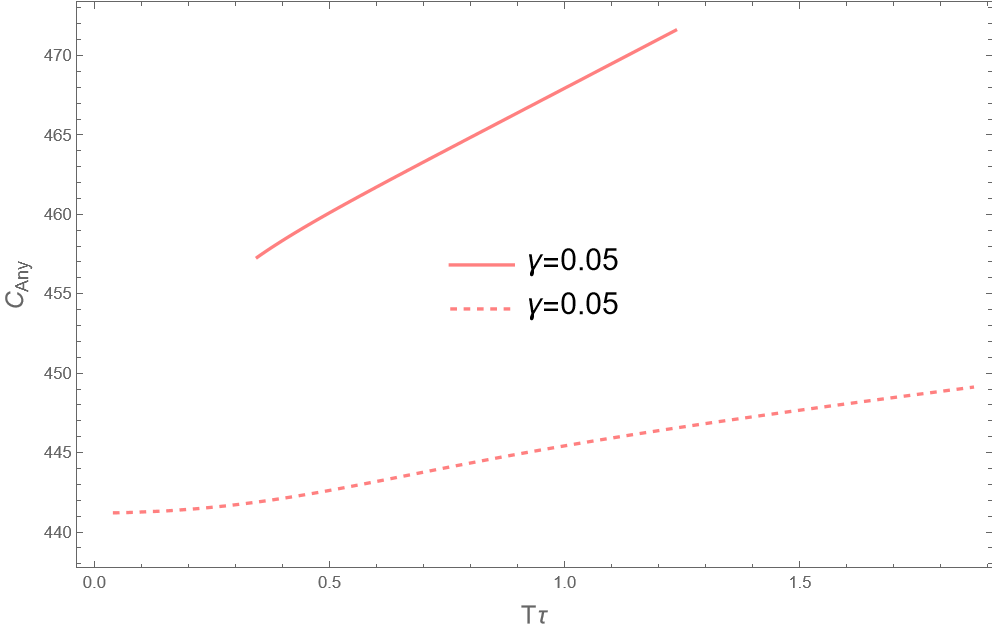} \label{fig:3f}}%
	\subfloat[]{\includegraphics[width=5.33cm]{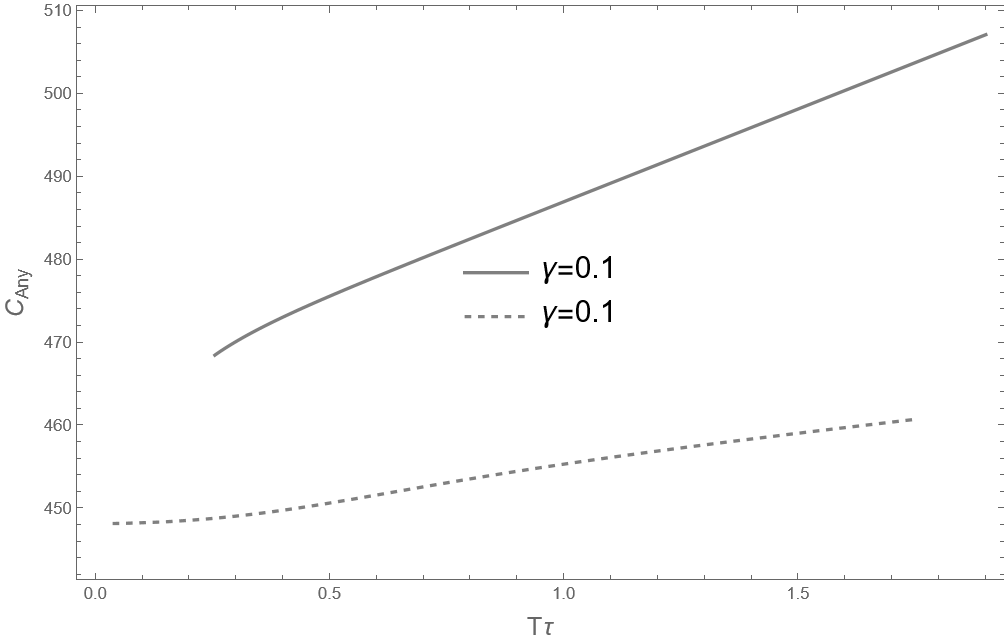} \label{fig:3g}}%
	\caption{(a) Effective potentials $U$ vs. radius $r$. (b--e) Boundary time multiplied by temperature vs. complexity time rate. (f, g) Complexity vs. boundary time multiplied by temperature for the 5-dimensional GB black brane with $r_{min}/r_+=0.5$, $\ell/r_+ = 1$, $\tilde{\alpha}/r_+^2 = 0.2$, and $\tilde{Q}/r_+^2 = 0.5$. The temperature $T$ and the outer horizon radius $r_+$ are evaluated in the limit $v\to\infty$. The dashed curves in the second and third rows mark the portions excluded by the non-maximality of the complexity.}\label{fig:3}%
\end{figure}

The blackening function of the charged AdS-Vaidya solution of Einstein theory, RN is \cite{rn}
\begin{align}\label{rn}
	f(r,v) = 1 - \frac{\tilde{M}(v)}{r^{d-2}} + \frac{r^2}{\ell^2} + \frac{\tilde{Q}(v)^2}{r^{2d-4}},
\end{align}
and that of GB theory is \cite{love}
\begin{align}\label{gb}
	f(r,v)=1+\frac{r^2}{2\tilde{\alpha}}\Big(1- \sqrt{1+4\tilde{\alpha}\big(\frac{\tilde{M}(v)}{r^d}-\frac{1}{\ell^2}-\frac{\tilde{Q}(v)^2}{r^{2d-2}}\big)}\Big),
\end{align}
where $ \tilde{M}(v) $ is the mass and $ \tilde{Q}(v) $ is the charge of the black branes, which increase with time. Here, $ \ell $ is the AdS radius, and $ \tilde{\alpha} $ is the coupling constant of the GB term in the action of the theory.

A typical choice for a smooth time change of parameters is:
\begin{align}\label{tran}
	\mathrm{f}_i(v) = \mathrm{f}_{i,1} + \frac{1}{2}(\mathrm{f}_{i,2} - \mathrm{f}_{i,1})\left[1 + \tanh\left(\frac{v - v_s}{\varepsilon}\right)\right],
\end{align}
where:
\begin{align}\label{param}
	&\mathrm{f}_i(v) = \left\lbrace \tilde{M}(v), \tilde{Q}(v)\right\rbrace.
\end{align}
Here $v_s$ is the shell time: the null coordinate of the center of the infalling shell, i.e.\ the location of the null hypersurface $v=v_s$ along which the perturbation is injected from the boundary. It fixes the boundary time at which the quench occurs, since the shell is emitted from $r\to\infty$ at
$t=v_s$. The constant $\varepsilon>0$ is the smoothing width, controlling the thickness of the shell in $v$.

Equation \eqref{tran} therefore interpolates between two stationary geometries: for $v \ll v_s$ one has $\mathrm{f}_i \to \mathrm{f}_{i,1}$ (the initial background, e.g.\ pure RN$\backslash$GB vacuum or a lighter black hole), while for $v \gg v_s$ one has $\mathrm{f}_i \to \mathrm{f}_{i,2}$ (the final black hole). In the thin-shell limit $\varepsilon \to 0$ the profile degenerates to a Heaviside step,
\begin{align}\label{heaviside}
	\mathrm{f}_i(v) \;\xrightarrow[\varepsilon \to 0]{}\;
	\mathrm{f}_{i,1} + \left(\mathrm{f}_{i,2}-\mathrm{f}_{i,1}\right)\Theta(v-v_s),
\end{align}
and the spacetime reduces to two exactly stationary regions glued across the null
shell $v=v_s$. Keeping $\varepsilon$ finite regulates this junction and is what makes the extremal-surface equations numerically tractable; unless stated otherwise we set $v_s=0$, so that $v<0$ and $v>0$ correspond to the pre- and post-quench regions respectively.

We consider a simplified version of the profile in \eqref{tran} where we set
$v_s = 0$ and
\begin{align}\label{paramv}
	&\mathrm{f}_{i,1}=0, \nonumber\\
	&\mathrm{f}_{i,2}=\left\lbrace  \tilde{M},  \tilde{Q}\right\rbrace,
\end{align}
so that a vacuum region is glued onto a one-sided black hole. With
\eqref{paramv}, each parameter follows
\begin{align}\label{paramtanh}
	\mathrm{f}_i(v) = \frac{\mathrm{f}_{i,2}}{2}
	\left[1 + \tanh\left(\frac{v}{\varepsilon}\right)\right],
\end{align}
and, since $f(r,v)$ in \eqref{rn} and \eqref{gb} depends on $v$ only through the $\mathrm{f}_i(v)$, the metric function smoothly varies between the two stationary solutions,
\begin{align}\label{bfs1}
	f(r,v) \;\longrightarrow\;
	\begin{cases}
		f_{vac}(r) = 1, & v \ll -\varepsilon, \\[4pt]
		f_{BH}(r), & v \gg +\varepsilon,
	\end{cases}
\end{align}
where $f_{BH}(r)$ is the same as \eqref{rn} and \eqref{gb} in each case. The transition is confined to
a shell of thickness $\sim\varepsilon$ centred on $v=0$; in the thin-shell limit
it degenerates to
\begin{align}\label{bfs2}
	f(r,v) \;\xrightarrow[\varepsilon\to0]{}\;
	1 + \left[f_{BH}(r)-1\right]\mathcal{H}(v),
\end{align}
i.e.\ the standard Vaidya gluing across the null shell $v=0$. We retain a small
but finite $\varepsilon$ throughout, both to regulate the discontinuity in the
extremal-surface equations and to model a quench of finite duration
$\Delta t \sim \varepsilon$ on the boundary.

The null energy condition (NEC) and weak energy condition (WEC) constrain this profile for our considered cases and are violated maximally for radii \cite{nec}
\begin{align}\label{nec}
	r^{d-2} < r^{d-2}_c:= \frac{\tilde{Q}(v)\,\partial_v\tilde{Q}(v)}{\partial_v\tilde{M}(v)}.
\end{align}
This means that the allowed radius range is equal to or greater than the critical radius $ r_c $.

Numerical examples of the effective potential and boundary time vs. $ \mathcal{C}_{Any} $ and its growth rate diagrams for RN black branes in 4d and 5d spacetimes are plotted in Fig.(\ref{fig:1}) and Fig.(\ref{fig:2}), respectively. In addition, these diagrams are plotted for the 5d GB case in Fig.(\ref{fig:3}). In the second line of all of the figures, it can be seen that for CAny cases, new branches appear in the complexity growth rate; these additional maxima in the potential are responsible for them \cite{callan, RN}.

In order to find the reliable branch, one need to consider validity the range of varying $ v $ that respects the NEC and WEC. Otherwise, these conditions can exclude the invalid range. This procedure may lead to a single branch. If the range of both branches observes the conditions, similarly to static instances, the maximal branch in the $\mathcal{C}_{Any}$ vs. boundary time diagrams determines the acceptable branch, due to complexity being defined as a maximum value \cite{rn}, like in our three examples. The corresponding branches are shown by solid lines, and the excluded portions by dashed lines in all three figures.  

It is worth noting that for the calculation of the complexity and its derivative with respect to the boundary time, one needs to extract the blackening function's vanishing points from the integration of \eqref{taup} which lead to a singularity. In the static (charged) case, there are two horizons and therefore, two such points. However, in the $ r $-constant cases, there are two such points for each varying time $ v $.

Therefore, compared to the complexity on $ v $-constant slices (static cases), there are one more level to find the valid range in $ r $-constant ones: first, NEC and WEC validity; second, maximal amounts of complexity. Obviously, the first one has higher priority.

Altogether, these analyses reveal that the effects of generalization on complexity and its time growth rate can occur on \textit{both} $ v $-constant and $ r $-constant surfaces in the AdS-Vaidya black hole background with smooth time change. In this regard, we refer to the complexity on $ r $-constant surfaces as ``doubled complexity" .

The results of this section motivate our subsequent discussions on comparing the behaviors of complexity generalization using the Weyl squared function of dynamical and static solutions in the boundary limit within FG coordinates.

%--------------------------------------------------------------------

\section{How complexity differs by injection of energy}\label{psec}

As discussed in the previous section, the static and dynamical solutions are not identical. In this section we show that their generically nonzero difference is sourced by the injection of energy in the Vaidya solution and encodes the resulting change in the dynamical response. Moreover, we should note that the material of this section is not restricted to the models we have studied.

It is worth mentioning that in a dynamical case, the change in the energy-momentum tensor is time-dependent, which can lead to altering the complexity growth rate. In the following we consider the complexity difference between the static and the dynamical solutions.

By the FG expansion of the induced metric determinant \cite{faraj}
\begin{align}\label{det}
\sqrt{h}=\frac{1}{2\rho^{\frac{d+1}{2}}}\Big(1+\frac{\rho G_{nn}}{2(d-2)}+\frac{(\rho-\rho_c)K^2}{2(d-1)}+...\Big),
\end{align}
where $ \rho $ is the FG radial coordinate, with the boundary at $ \rho=0 $, inserting the FG expansion of the Weyl squared tensor and \eqref{det} in \eqref{maxv} one finds (see the Appendix \ref{fgex} for details) \footnote{We should note that we conduct the following analysis for $d=4+1$, but it can be generalized for an arbitrary dimension.}
\begin{align}\label{c14}
\mathcal C_{Any}=\int d^{d-1}\sigma d\rho \frac{1}{2\rho^{\frac{d+1}{2}}}\Big(1+\rho C_1+\rho^2 C_2+\rho^3 C_3+ \rho^4 C_4 +...\Big)
\end{align}
Now, subtracting the complexity of the dynamical solution and the static solution one finds
\begin{align}\label{subt}
\mathcal C_{Any}^d-\mathcal C_{Any}^s=\int d^{d-1}\sigma d\rho \frac{1}{2\rho^{\frac{d+1}{2}}}\Big(\rho (C_1^d-C_1^s)+\rho^2 (C_2^d-C_2^s)+...\Big)
\end{align}
Because of the fact that both solutions are asymptotic AdS and identical at the boundary $g^{(0)}_d=g^{(0)}_s$, then $C_1=C_2=C_3=C_4$ are identical in either solutions (they all are written in terms of the boundary metric $g^{(0)}$, $g^{(2)}$, $X^{(0)}$ and $X^{(1)}$, the embedding at  zeroth and first order\footnote{See Appendix \ref{fgex} for details.}). Hence, \eqref{subt} at the leading non-vanishing order is given by
\begin{align}\label{difference}
\Delta \mathcal C_{Any}=\mathcal C_{Any}^d-\mathcal C_{Any}^s=\int d^{d-1}\sigma d\rho \frac{1}{2\rho^{\frac{d+1}{2}}}\Big(\rho^5 (C_5^d-C_5^s)+\mathcal O(\rho^6)\Big)
\end{align}
The subtraction of $C_5$ is given (see Appendix \ref{fgex})
\begin{align}
C_5^d-C_5^s=&\Big(\frac{G^{(4)}}{2(d-2)}+\frac{K^2_{(4)}}{2(d-1)}\Big)|^d-\Big(\frac{G^{(4)}}{2(d-2)}+\frac{K^2_{(4)}}{2(d-1)}\Big)|^s
\end{align}
where 
\begin{align}
&\frac{2L^2}{d(d-1)}G^{(4)}=g_{\mu\nu}^{(4)}n^{(0)\mu}n^{(0)\nu}+2g_{\mu\nu}^{(2)}n^{(0)\mu}n^{(1)\nu}+g_{\mu\nu}^{(0)}n^{(1)\mu}n^{(1)\nu}+2g_{\mu\nu}^{(0)}n^{(0)\mu}n^{(2)\nu}
,\\
&K_{(4)}^2=\left(h^{(2)\mu\nu}N_{\mu\nu}^{(0)}+h^{(0)\mu\nu}N_{\mu\nu}^{(2)}\right)^2+2h^{(0)\alpha\beta}N_{\alpha\beta}^{(0)}\left(h^{(4)\mu\nu}N_{\mu\nu}^{(0)}+h^{(2)\mu\nu}N_{\mu\nu}^{(2)}+h^{(0)\mu\nu}N_{\mu\nu}^{(4)}\right).
\end{align}
Due to the point that $g^{(4)}\sim <T>$ and $<T>_s$ is constant \cite{sken} \footnote{It is worth mentioning that in homogeneous spaces, there is spacial symmetry and as a consequence, the Lie derivative of the boundary metric is vanished, which means that $\partial_{i}<T_{ab}>=0$; in addition, in static solution $\partial_t<T_{ab}>=0$. This means that $<T_{ab}>$ is a constant of time and spacial coordinates.}, and the fact that the solutions \eqref{rn} and \eqref{gb} are asymptotically AdS, the subtraction leads to
\begin{align}\label{c5}
C_5^d-C_5^s=&\frac{1}{2(d-2)}\Big(G^{(4)}_d-G^{(4)}_s+K^2_{(4),d}-K^2_{(4),s}\Big),\\
&=\frac{1}{2(d-2)}\Bigg(\frac{d(d-1)}{L^2}n^{(0)\mu}n^{(0)\nu}(T_{\mu\nu,d}-T_{\mu\nu,s})+2\left(h^{(0)\alpha\beta}N_{\alpha\beta}^{(0)}N_{\mu\nu}^{(0)}\right)(T_d^{\mu\nu}-T_s^{\mu\nu})\nonumber\\
&-\left(h^{(0)\alpha\beta}N_{\alpha\beta}^{(0)}h^{(0)\mu\nu}n_\lambda^{(0)}\right)\Big( g^{(0)\lambda l}\left(\nabla_\mu T_{l\nu}+\nabla_\nu T_{\mu l}-\nabla_l T_{\mu\nu}\right)_d\nonumber\\
&-(\nabla_\mu g^{(0)}_{l\nu}+\nabla_\nu g^{(0)}_{\mu l}-\nabla_l g^{(0)}_{\mu\nu})(T_d^{\lambda l}-T_s^{\lambda l})\Big)\Bigg)
\end{align}
Eq. \eqref{c5} shows that the subtraction of the complexity for the static and Vaidya solutions at the leading order contains terms proportional to the energy-momentum tensor, its derivatives and some terms containing the boundary metric $g^{(0)}$. We now consider \eqref{c5} before, during and after the quench:

\textbf{Before the quench $(v\ll v_s)$: $T_s=T_0, T_d=0, \nabla T_d=0$}

Before the quench, the system is in thermal equilibrium. The complexity difference arises solely from the static contribution $T_s$, which is independent of the dynamical response.

\textbf{After the quench $(v\gg v_s)$: $T_s=T_0, T_d=T_0, \nabla T_d=0$}

Eq. \eqref{c5} would vanish.

\textbf{During the quench $(v\approx v_s)$: $T_s=T_0, T_d=T_0\Theta(v), \nabla T_d\sim \mathcal T(v)$}

$\Theta(v)$ would be the Heaviside function in the sharp quench and in the smooth quench would be $1+tanh(\frac{v-v_s}{\varepsilon})$, where $\varepsilon$ characterized the smoothness. Respectively, $\mathcal T\sim \delta(v)$ and $\mathcal T \sim sech^2(\frac{v-v_s}{\varepsilon})$. In the proceeding section we consider the sharp and the smooth profile.
%--------------------------
\subsection{Thin-shell collapse: $\mathcal T\sim \delta(v)$}
In the region near the shell collapsing along the null surface $v=v_s$, there is $T_d=T_s$, then terms containing $\Delta T=T_d-T_s$, vanish and only $\nabla T$ survives. Because of the fact that we supposed the energy momentum tensor in the Vaidya solution is proportional to Heaviside function and its derivative leads to delta Dirac function where its integration is not vanished $\int d^{d-1}d\rho \delta(v(\rho))\neq 0$. Then \eqref{c5} in that region simplifies as
\begin{align}\label{regi}
C_5^d-C_5^s=&\frac{-1}{2(d-2)}h^{(0)\alpha\beta}N_{\alpha\beta}^{(0)}h^{(0)\mu\nu}n_\lambda^{(0)} g^{(0)\lambda l}\left(\nabla_\mu T_{l\nu}+\nabla_\nu T_{\mu l}-\nabla_l T_{\mu\nu}\right)_d
\end{align}
In linear response theory we know how the expectation of the energy momentum tensor is changed
\begin{align}\label{deltatij}
\delta<T_{ij}(t)>=\int dt' G^R_{ij,kl}(t, t') J^{kl}(t')
\end{align}
where $G^R$ is the retarded green's function. Introducing average and relative time, respectively as \footnote{We should note that for the time-dependent quench, we use the generalized response theory version of the dispersion relations rather than assuming time translation invariance of the perturbed state.}
\begin{align}
&A=\frac{t+t'}{2},\\
&B=t-t',
\end{align}

Then, \eqref{deltatij} would be
\begin{align}\label{fdeltat}
\delta <T_{ij}(t)>=\int dB~ G^R_{ij,kl}(t-\frac{B}{2},B)J^{kl}(t-B).
\end{align}
the retarded Green's function can then be written as $G^R_{ij,kl}(t-\frac{B}{2},B)=G^R_{ij,kl}(A+\frac{B}{2},A-\frac{B}{2})$. By the inverse Wigner transformation \cite{kk},
\begin{align}
G^R_{ij,kl}(A+\frac{B}{2},A-\frac{B}{2})=\int \frac{d\omega}{2\pi} e^{-i\omega B}G^R_{ij,kl}(\omega,A),
\end{align}
where $\omega$ is the frequency in the Wigner transformation. By inserting the inverse Wigner transform of the Green's function into \eqref{fdeltat},
\begin{align}
\delta<T_{ij}(t)>= \int dB~\frac{d\omega}{2\pi} e^{-i\omega B}G^R_{ij,kl}(\omega,A)J^{kl}(t-B),
\end{align}

The spectral function is written as follows \cite{specfun}
\begin{align}
\rho_{ij,kl}(\omega,A)=-2\operatorname{Im} G^R_{ij,kl}(\omega,A)
\end{align}
So time derivative of the expectation value of the energy-momentum tensor $\partial_t<T_{ij}>$ is therefore\footnote{By derivative some boundary terms appear as $G^R_{boundary}$ in the integral. We have omitted the equal-time boundary contribution $G^R(t,t')J(t')$, which is a local contact term arising from differentiating the upper limit of the causal response integral. We adopt the renormalization prescription in which such local terms are subtracted. Since these terms have no absorptive part, this subtraction does not modify the spectral density or the nonlocal causal response encoded by it.},
\begin{align}\label{parten}
\partial_t<T_{ij}>&=\frac{1}{2}\partial_A<T_{ij}>+\partial_B<T_{ij}>\\
&=\int dB~\frac{d\omega}{2\pi} (\frac{1}{2}\partial_A+\partial_B)\Big(e^{-i\omega B} G_{ij,kl}^R(\omega,A) J^{kl}(t-B)\Big).
\end{align}
Green's functions could be expanded in terms of the imaginary and real parts as
\begin{align}
G^R(\omega,A)&=\operatorname{Re}G^R(\omega,A)+i~\operatorname{Im}G^R(\omega,A)
\end{align}
For a retarded Green's function $ G^R(\omega,A) $, which is analytic in the upper half-plane of complex frequency $ \omega $, the Kramers-Kronig (KK) relations state \cite{kk} \footnote{We should note that the KK relations in the linear response theory are based on causality and time invariance, where in the presence of perturbation the time invariance is ruined. However, the causality allows to generalize the KK relations to use them in the states of far from equilibrium (see Appendix \ref{green} for details). Although, \cite{kk} applies this method to the susceptibility of the system, the susceptibility is the Fourier transform of the Green's function.}
\begin{align} 
&\operatorname{Re}G^R(\omega,A) = \frac{2}{\pi} \mathcal{P} \int_{-\infty}^{\infty} \frac{\omega'\operatorname{Im}G^R(\omega',A)}{\omega'^2 - \omega^2} d\omega'\\  
&\operatorname{Im}G^R(\omega,A) = -\frac{2\omega}{\pi} \mathcal{P} \int_{-\infty}^{\infty} \frac{\operatorname{Re}G^R(\omega',A)}{\omega'^2 - \omega^2} d\omega'
\end{align} 
where $ \mathcal{P} $ denotes the Cauchy principal value. These relations imply that the real and imaginary parts of $ G^R(\omega,A) $ are not independent, one determines the other via an integral transform. Then, all terms in the right hand side of \eqref{parten} could be rewritten in terms of the imaginary part of the Green's function and as a consequence in terms of the stress tensor spectral function of the boundary theory. Resultantly, \eqref{parten} could be written in terms of the spectral function as follows \footnote{Here the KK transform converts the real part of the retarded correlator into an integral over its imaginary part, so the response can be represented as a functional of the spectral density, with a correspondingly transformed frequency kernel.}
\begin{align}
\partial_t<T_{ij}>&\sim \int dB~d\omega  (\frac{1}{2}\partial_A+\partial_B)\Big[e^{-i\omega B} J^{kl}(t-B)\Big(\frac{i ~\rho_{ij,kl}(\omega,A)}{2}+\frac{1}{\pi}\int \frac{\omega' \rho_{ij,kl}(\omega',A)}{\omega'^2-\omega^2}d\omega'\Big)\Big],\\
&:=\int d\Omega \mathcal K^{kl}(A,B,\omega,\omega') \rho_{ij,kl}(\omega',A),\label{gr}
\end{align}
where $d\Omega=dB~d\omega$ and $\mathcal K^{kl}$ is the kernel of the integral
\begin{align}
\mathcal K^{kl}(A,B,\omega,\omega')=(\frac{1}{2}\partial_A+\partial_B)\Big[e^{-i\omega B} J^{kl}(t-B)\Big(\frac{i \delta(\omega-\omega')}{2}+\frac{1}{\pi}\int \frac{\omega' }{\omega'^2-\omega^2}d\omega'\Big)\Big]
\end{align}

Substituting \eqref{gr} into \eqref{difference} leads to
\begin{align}\label{cti}
\Delta \mathcal{C}_{Any}\sim \int \sqrt{h}d^{d-1}\sigma d\rho \rho^{\frac{-d+9}{2}} \mathcal H^{ij}(g^{(0)})\int d\Omega \mathcal K^{kl}(A,B,\omega,\omega')\rho_{ij,kl}(\omega',A)
\end{align}
where $\mathcal H^{ij}(g^{(0)})$ stands for the terms in \eqref{regi}. As \eqref{cti} reveals, the difference of the complexity for the static and the Vaidya solutions is associated with the dynamical response generated during the Vaidya collapse. The FG expansion shows that the corresponding contribution to the complexity difference is controlled by the time-dependent boundary stress tensor. Within linear response, this stress tensor response is determined by the retarded Green's function and can consequently be expressed in terms of the corresponding the stress tensor spectral function of the boundary theory.

%----------------------------------------------------------------------
\section{Conclusion}

\label{sec:conclusion}

In this work, we have investigated the dynamics of generalized holographic complexity in Vaidya 4- and 5-dimensional RN and 5-dimensional GB black branes. We focused on the complexity=anything prescription with a Weyl squared correction and compared the complexity dynamics of Vaidya geometries with those of static ones. Our study also includes a complementary near-boundary analysis based on the Fefferman-Graham expansion.

For the explicit bulk solution, the complexity growth rate in the CAny proposal, compared to CV, can generate additional branches. Recent work has studied these effects in static and Vaidya solutions with instantaneous shock waves. However, the Vaidya case with a smooth time variation range has been overlooked. In this paper, we have shown that for such a case, the dynamical behavior of volume complexity and its generalization to CAny on $ r $-constant surfaces is similar to that on $ v $-constant surfaces. This "doubled complexity" behavior, similar to the static case, also contains additional branches for CAny as compared to the CV case. Thus, the Weyl squared term, which we use in the generalization function, provides information about the bulk geometry that is not captured by the standard volume functional. The explicit location and detailed behavior of these branches, as well as their dependence on the model parameters, should be regarded as properties of the particular holographic realization studied here.

We have then considered the difference between the generalized complexities of the dynamical and static geometries,
\begin{equation}
\Delta \mathcal C_{Any}=\mathcal C_{Any}^{d}-\mathcal C_{Any}^{s},
\end{equation}
in the asymptotic AdS region. In this part of the analysis, we have used the Fefferman-Graham expansion of the bulk metric together with the expansions of the extremal embedding, induced metric, normal vector, and extrinsic curvature. A central result is that the first four coefficients in the complexity expansion coincide for the static and dynamical geometries. This equality follows from the fact that these coefficients are determined by the common asymptotic boundary data, together with the corresponding near-boundary bulk and embedding expansions. The first nontrivial contribution therefore appears only at the subsequent order.

The resulting leading contribution to the complexity difference is governed by the boundary stress tensor and its derivatives. For the Vaidya collapse, the two stress tensors coincide in the region of the thin shell in the sense relevant for the subtraction, so that the contribution proportional to $\Delta T_{\mu\nu}$ vanishes there, while the derivative of the dynamical stress tensor remains. The nonvanishing contribution to the complexity subtraction is therefore directly associated with the time-dependent response produced by energy injection. This gives a geometric interpretation of the difference between the static and dynamical complexities. We have further related this result to boundary response theory. In linear response, the time-dependent stress tensor is determined by the retarded Green's function. Using the causal dispersion formulas, the dynamical stress tensor contribution entering the complexity difference can then be written in terms of the spectral function. 

It is useful to distinguish the universal content of these results from the properties of the particular bulk model. The additional complexity branches and their detailed dependence on the bulk parameters are properties of the specific geometric realization used for the explicit calculation. In contrast, the cancellation of the first four terms in the FG expansion, the appearance of the boundary stress tensor in the first nonvanishing contribution, and the response theory interpretation follow from the asymptotic structure and do not require the detailed microscopic realization of the bulk matter sector. The specific model therefore serves as a concrete computational realization in which these general features can be exhibited explicitly.

Our results suggest that generalized holographic complexity can provide a geometric probe of dynamical response in strongly coupled quantum systems. In particular, the Weyl squared contribution makes the complexity sensitive to dynamical bulk structures invisible to the standard CV functional, while the FG subtraction isolates the part of the complexity controlled by the time-dependent boundary stress tensor. It would be interesting to investigate whether analogous relations persist for other choices of the generalized complexity functional and for more general nonequilibrium processes. In particular, different curvature invariants may probe different combinations of boundary response functions, offering a possible route to a broader classification of the spectral information encoded in generalized holographic complexity.
%--------------------------------------------------------------------------

\section*{Acknowledgment}
The authors would like to express their gratitude to Shahrokh Parvizi for useful discussions and for reading the manuscript. In addition, they extend their deepest condolences to the families and loved ones of the students who lost their lives at the school in Minab, Iran, and to the Iranian people on that Day. This loss is a profound tragedy, and they honor their memory with respect and sorrow. May their souls rest in peace.

%-------------------------------------------------------------------------------
\appendix
\renewcommand\theequation{\thesection-\arabic{equation}} 
\setcounter{equation}{0}

%-------------------------------------------------------------

\section{The generalized Kramers-Kronig relations}\label{green}
In this section we use the Kramers-Kronig (KK) relations and the properties of the Green's function integral to show that the integral of the multiplication of the cross-terms (real part into the imaginary part) would be turn into the imaginary part of the Green's function.

The generalized Kramers-Kronig (KK) relations follow from the causality of the
retarded Green's function. In the nonequilibrium setting considered here,
the retarded correlator depends on two independent times. Introducing the
average and relative times,
\begin{equation}
A=\frac{t+t'}{2},\qquad B=t-t',
\end{equation}
we define its Wigner transform with respect to the relative time,
\begin{equation}
G^R(\omega,A)
=
\int_0^\infty dB\,e^{i\omega B}
G^R\left(A+\frac{B}{2},A-\frac{B}{2}\right).
\end{equation}
For each fixed average time $A$, causality implies that
$G^R(\omega,A)$ is analytic in the upper half of the complex $\omega$-plane. Consequently, its real and imaginary parts are related by the generalized
KK relations \cite{kk},
\begin{equation}
\operatorname{Re}G^R(\omega,A)
=
\frac{1}{\pi}\mathcal{P}
\int_{-\infty}^{\infty}d\omega'\,
\frac{\operatorname{Im}G^R(\omega',A)}
{\omega'-\omega},
\label{KK-real}
\end{equation}
and
\begin{equation}
\operatorname{Im}G^R(\omega,A)
=
-\frac{1}{\pi}\mathcal{P}
\int_{-\infty}^{\infty}d\omega'\,
\frac{\operatorname{Re}G^R(\omega',A)}
{\omega'-\omega},
\end{equation}
where $ \mathcal{P} $ denotes the Cauchy principal value. These relations imply that the real and imaginary parts of $ G^R(\omega,A) $ are not independent, one determines the other via an integral transform.

%------------------------------------------------------------------------
\section{The FG expansion of $G_{nn}$, $K^2$ and complexity functional}\label{fgex}
The purpose of this appendix is to show explicitly that the equality of the first four coefficients in \eqref{c14} follows after simultaneously expanding the bulk metric, the extremal embedding, the normal vector, the induced metric, and the extrinsic curvature. In this section the detail of the FG expansion of $G_{nn}$, the extrinsic curvature and the complexity functional are computed.
%---------------------------------------------------------------------------
\subsection{FG expansion of $G_{nn}$ and $K^2$}
We begin with the FG expansion of an asymptotically AdS metric,
\begin{equation}
ds^2=\frac{L^2}{4\rho^2}d\rho^2+\frac{L^2}{\rho}g_{ij}(x\rho)\,dx^idx^j ,
\end{equation}
where
\begin{equation}
g_{ij}(x,\rho)=g_{ij}^{(0)}+\rho g_{ij}^{(2)}+\rho^2g_{ij}^{(4)}+\mathcal O(\rho^3).
\end{equation}
The FG expansion of the Christoffel symbol would be
\begin{align}
\Gamma^i_{jk}=\Gamma^{(0)i}_{jk}+\rho \Gamma^{(2)i}_{jk}+\rho^2 \Gamma^{(4)i}_{jk}+\cdots ,
\end{align}
where
\begin{align}\label{gamm}
&\Gamma^{(0)i}_{jk}=\frac12 g^{(0)il}\left(\nabla_j g^{(0)}_{lk}+\nabla_k g^{(0)}_{jl}-\nabla_l g^{(0)}_{jk}\right),\\
&\Gamma^{(2)i}_{jk}=\frac12 g^{(0)il}\left(\nabla_j g^{(2)}_{lk}+\nabla_k g^{(2)}_{jl}-\nabla_l g^{(2)}_{jk}\right)-\frac12 g^{(2)il}\left(\nabla_j g^{(0)}_{lk}+\nabla_k g^{(0)}_{jl}-\nabla_l g^{(0)}_{jk}\right),\\
&\Gamma^{(4)i}_{jk}=\frac12 g^{(0)il}\left(\nabla_j g^{(4)}_{lk}+\nabla_k g^{(4)}_{jl}-\nabla_l g^{(4)}_{jk}\right)-\frac12 g^{(2)il}\left(\nabla_j g^{(2)}_{lk}+\nabla_k g^{(2)}_{jl}-\nabla_l g^{(2)}_{jk}\right)\nonumber\\
&-\frac12 g^{(4)il}\left(\nabla_j g^{(0)}_{lk}+\nabla_k g^{(0)}_{jl}-\nabla_l g^{(0)}_{jk}\right)+\frac12 g^{(2)im}g^{(2)l}_m\left(\nabla_j g^{(0)}_{lk}+\nabla_k g^{(0)}_{jl}-\nabla_l g^{(0)}_{jk}\right)
\end{align}
%---------------------------------
\subsection{The FG expansion of $K^2$}
For the extremal hypersurface parameterized by
\begin{equation}
X^{\mu}=\left(t(\rho),\rho,x^{i}\right),
\end{equation}
we expand it as in \cite{comment}
\begin{align}
X^{\mu}=X_{(0)}^\mu+\rho X_{(1)}^\mu+\mathcal O(\rho^2).
\end{align}
Similar to \cite{comment} we choose the gauge $X^\mu=(t,\rho,x^i)$. By the fact that the unit vector satisfies orthogonality and unitness
\begin{align}
&g_{\mu\mu}n^\mu n^\nu=-1,\\
&g_{\mu\nu}n^\mu e_i^\nu,
\end{align}
where $e_i^\nu=\partial_i X^\nu=\sum \rho^ne_i^{(n)\nu}$, is the tangent vector one finds
\begin{align}
n^\mu=n^{(0)\mu}+\rho n^{(1)\mu}+\cdots.
\end{align}
The orders of $n$ are given by the orthogonality condition
\begin{align}\label{nvector}
&g_{\mu\nu}^{(0)}n^{(0)\mu}e_i^{(0)\nu}=0,\\
&g_{\mu\nu}^{(0)}n^{(1)\mu}e_i^{(0)\nu}+g_{\mu\nu}^{(0)}n^{(0)\mu}e_i^{(1)\nu}+g_{\mu\nu}^{(2)}n^{(0)\mu}e_i^{(0)\nu}=0,\\
&g_{\mu\nu}^{(0)}n^{(2)\mu}e_i^{(0)\nu}+g_{\mu\nu}^{(0)}n^{(1)\mu}e_i^{(1)\nu}+g_{\mu\nu}^{(0)}n^{(0)\mu}e_i^{(2)\nu}\nonumber\\
&+g_{\mu\nu}^{(2)}n^{(1)\mu}e_i^{(0)\nu}+g_{\mu\nu}^{(2)}n^{(0)\mu}e_i^{(1)\nu}+g_{\mu\nu}^{(4)}n^{(0)\mu}e_i^{(0)\nu}=0.\label{orth}
\end{align}
It has been shown that $X_{(1)}^\mu \sim Kn^\mu$ is written in terms of $X_{(0)}^\mu$, \cite{comment} \footnote{We should note that in \cite{comment}, $X^\mu$ has been considered in CV; however, due to the fact that the FG expansion of the Weyl squared tensor starts from $\rho^2$, $X^\mu$ and $n^\mu$ expansions are the same in CV and the generalized complexity up to the first order.} and therefore the embedding corrections do not contribute up to $\mathcal{O}(\rho^2)$. The tangent vectors are
\begin{equation}
e_{\rho}^{\mu}=\left(t',1,0,\cdots\right),\qquad e_{i}^{\mu}=\delta_{i}^{\mu}.
\end{equation}
The equation for the embedding in terms of \eqref{maxv} would be
\begin{equation}
\partial_i \left(\sqrt{h}h^{ij}g_{\mu\nu}\partial_j X^\nu  \mathcal{F}_1\right)-\frac12 \mathcal{F}_1\sqrt{h}h^{ij}\partial_\mu g_{\alpha \beta}\partial_i X^\alpha \partial_j X^\beta - \sqrt{h}\partial_\mu \mathcal{F}_1=0
\end{equation}

The extrinsic curvature is defined by
\begin{equation}
K_{\mu\nu}=h_{\mu}^{\ \alpha}h_{\nu}^{\ \beta}\nabla_{\alpha}n_{\beta}.
\end{equation}
we define $\nabla_\mu n_\nu:=N_{\mu\nu}$, where
\begin{equation}
N_{\mu\nu}=N_{\mu\nu}^{(0)}+\rho N_{\mu\nu}^{(2)}+\rho^2N_{\mu\nu}^{(4)}+\cdots.
\end{equation}
These terms are given by
\begin{align}\label{N}
&N_{\mu\nu}^{(0)}=\partial_\mu n_\nu^{(0)}-\Gamma_{\mu\nu}^{(0)\lambda}n_\lambda^{(0)},\\
&N_{\mu\nu}^{(2)}=\partial_\mu n_\nu^{(1)}-\Gamma_{\mu\nu}^{(0)\lambda}n_\lambda^{(1)}-\Gamma_{\mu\nu}^{(2)\lambda}n_\lambda^{(0)},\\
&N_{\mu\nu}^{(4)}=\partial_\mu n_\nu^{(2)}-\Gamma_{\mu\nu}^{(0)\lambda}n_\lambda^{(2)}-\Gamma_{\mu\nu}^{(2)\lambda}n_\lambda^{(1)}-\Gamma_{\mu\nu}^{(4)\lambda}n_\lambda^{(0)}.
\end{align}
By defining $h_{\mu\nu}=g_{\mu\nu}-n_\mu n_\nu$, the FG expansion of the induced metric is found
\begin{align}\label{h}
&h_{\mu\nu}=h_{\mu\nu}^{(0)}+\rho h_{\mu\nu}^{(2)}+\rho^2 h_{\mu\nu}^{(4)}+\cdots,\\
&h_{\mu\nu}^{(0)}=g_{\mu\nu}^{(0)}-n_\mu^{(0)}n_\nu^{(0)},\\
&h_{\mu\nu}^{(2)}=g_{\mu\nu}^{(2)}-n_\mu^{(1)}n_\nu^{(0)}-n_\mu^{(0)}n_\nu^{(1)},\\
&h_{\mu\nu}^{(4)}=g_{\mu\nu}^{(4)}-n_\mu^{(2)}n_\nu^{(0)}-n_\mu^{(0)}n_\nu^{(2)}-n_\mu^{(1)}n_\nu^{(1)}.
\end{align}
Using the above expansions, the the extrinsic curvature would be
\begin{align}
&K_{\mu\nu}=K_{\mu\nu}^{(0)}+\rho K_{\mu\nu}^{(2)}+\rho^2K_{\mu\nu}^{(4)}+\cdots,\\
&K_{\mu\nu}^{(0)}=h_\mu^{(0)\alpha} h_\nu^{(0)\beta}N_{\alpha \beta}^{(0)},\\
&K_{\mu\nu}^{(2)}=h_\mu^{(2)\alpha} h_\nu^{(0)\beta}N_{\alpha \beta}^{(0)}+h_\mu^{(0)\alpha} h_\nu^{(2)\beta}N_{\alpha \beta}^{(0)}+h_\mu^{(0)\alpha} h_\nu^{(0)\beta}N_{\alpha \beta}^{(2)},\\
&K_{\mu\nu}^{(4)}=h_\mu^{(4)\alpha} h_\nu^{(0)\beta}N_{\alpha \beta}^{(0)}+h_\mu^{(0)\alpha} h_\nu^{(4)\beta}N_{\alpha \beta}^{(0)}+h_\mu^{(2)\alpha} h_\nu^{(2)\beta}N_{\alpha \beta}^{(0)}\nonumber\\
&+h_\mu^{(2)\alpha} h_\nu^{(0)\beta}N_{\alpha \beta}^{(2)}+h_\mu^{(0)\alpha} h_\nu^{(2)\beta}N_{\alpha \beta}^{(2)}+h_\mu^{(0)\alpha} h_\nu^{(0)\beta}N_{\alpha \beta}^{(4)}.
\end{align}
The trace of the extrinsic curvature is written as
\begin{align}
&g^{\mu\nu}K_{\mu\nu}=K^{(0)}+\rho K^{(2)}+\rho^2K^{(4)}+\cdots,\\
&K^{(0)}=h^{(0)\mu\nu}N_{\mu\nu}^{(0)},\\
&K^{(2)}=h^{(2)\mu\nu}N_{\mu\nu}^{(0)}+h^{(0)\mu\nu}N_{\mu\nu}^{(2)},\\
&K^{(4)}=h^{(4)\mu\nu}N_{\mu\nu}^{(0)}+h^{(2)\mu\nu}N_{\mu\nu}^{(2)}+h^{(0)\mu\nu}N_{\mu\nu}^{(4)}.
\end{align}
Finally,
\begin{align}
&K^2=K_{(0)}^2+\rho K_{(2)}^2+\rho^2 K_{(4)}^2+\cdots,\\
&K_{(0)}^2=\left(h^{(0)\mu\nu}N_{\mu\nu}^{(0)}\right)^2,\\
&K_{(2)}^2=2\left(h^{(0)\alpha\beta}N_{\alpha\beta}^{(0)}\right)\left(h^{(2)\mu\nu}N_{\mu\nu}^{(0)}+h^{(0)\mu\nu}N_{\mu\nu}^{(2)}\right),\\
&K_{(4)}^2=\left(h^{(2)\mu\nu}N_{\mu\nu}^{(0)}+h^{(0)\mu\nu}N_{\mu\nu}^{(2)}\right)^2\nonumber\\
&+2h^{(0)\alpha\beta}N_{\alpha\beta}^{(0)}\left(h^{(4)\mu\nu}N_{\mu\nu}^{(0)}+h^{(2)\mu\nu}N_{\mu\nu}^{(2)}+h^{(0)\mu\nu}N_{\mu\nu}^{(4)}\right).
\end{align}
To compare $K^2$ in the static and the Vaidya solutions, we should note that $g^{(0)}$ is identical in either solutions and as a consequence $g^{(2)}$ (which is written in terms of $g^{(0)}$). In addition, \eqref{nvector} shows that the zeroth order of the normal vector $n^{(0)}$ and the first order $n^{(1)}$ are identical in either solutions because they are written in terms of $g^{(0)}$, $g^{(2)}$, $X^{(0)}$ and $X^{(1)}$. Then, 
\begin{align}
&K^2_{(4),v}-K^2_{(4),s}=\nonumber\\
&2\Bigg(h^{(0)\alpha\beta}N_{\alpha\beta}^{(0)}\left(h^{(4)\mu\nu}N_{\mu\nu}^{(0)}+h^{(0)\mu\nu}N_{\mu\nu}^{(4)}\right)\Bigg)_v-2\Bigg(h^{(0)\alpha\beta}N_{\alpha\beta}^{(0)}\left(h^{(4)\mu\nu}N_{\mu\nu}^{(0)}+h^{(0)\mu\nu}N_{\mu\nu}^{(4)}\right)\Bigg)_s
\end{align}
By use of \eqref{N} and \eqref{h}, the above subtraction would be
\begin{align}
&K^2_{(4),v}-K^2_{(4),s}=2\left(h^{(0)\alpha\beta}N_{\alpha\beta}^{(0)}N_{\mu\nu}^{(0)}\right)(g_v^{(4)\mu\nu}-g_s^{(4)\mu\nu})\nonumber\\
&-2\left(h^{(0)\alpha\beta}N_{\alpha\beta}^{(0)}h^{(0)\mu\nu}n_\lambda^{(0)}\right)(\Gamma_{\mu\nu,v}^{(4)\lambda}-\Gamma_{\mu\nu,s}^{(4)\lambda})
\end{align}
Using \eqref{gamm},
\begin{align}
&\Gamma^{(4)\lambda}_{\mu\nu,v}-\Gamma^{(4)\lambda}_{\mu\nu,s}=\frac12 g^{(0)\lambda l}\Big(\left(\nabla_\mu g^{(4)}_{l\nu}+\nabla_\nu g^{(4)}_{\mu l}-\nabla_l g^{(4)}_{\mu\nu}\right)_v-\left(\nabla_\mu g^{(4)}_{l\nu}+\nabla_\nu g^{(4)}_{\mu l}-\nabla_l g^{(4)}_{\mu\nu}\right)_s\Big)\nonumber\\
&-\frac12 \left(\nabla_\mu g^{(0)}_{l\nu}+\nabla_\nu g^{(0)}_{\mu l}-\nabla_l g^{(0)}_{\mu\nu}\right)(g_v^{(4)\lambda l}-g_s^{(4)\lambda l})
\end{align}
Due to the point that $g^{(4)}\sim <T>:=T$ and $T_s=cte$, the subtraction leads to
\begin{align}
&K^2_{(4),v}-K^2_{(4),s}=2\left(h^{(0)\alpha\beta}N_{\alpha\beta}^{(0)}N_{\mu\nu}^{(0)}\right)(T_v^{\mu\nu}-T_s^{\mu\nu})\nonumber\\
&-2\left(h^{(0)\alpha\beta}N_{\alpha\beta}^{(0)}h^{(0)\mu\nu}n_\lambda^{(0)}\right)\Bigg(\frac12 g^{(0)\lambda l}\left(\nabla_\mu T_{l\nu}+\nabla_\nu T_{\mu l}-\nabla_l T_{\mu\nu}\right)_v\nonumber\\
&-\frac12 (\nabla_\mu g^{(0)}_{l\nu}+\nabla_\nu g^{(0)}_{\mu l}-\nabla_l g^{(0)}_{\mu\nu})(T_v^{\lambda l}-T_s^{\lambda l})\Bigg)
\end{align}
%----------------------------------------------
\subsubsection{Expansion of $G_{nn}$}
The Einstein tensor is
\begin{equation}
G_{\mu\nu}=R_{\mu\nu}-\frac12 g_{\mu\nu}R.
\end{equation}
Projecting along the unit normal,
\begin{equation}
G_{nn}=n^\mu n^\nu G_{\mu\nu}.
\end{equation}
Expanding around AdS in FG,
\begin{align}
&G_{nn}=G^{(0)}+\rho G^{(2)}+\rho^2 G^{(4)}+\cdots,\\
&\frac{2L^2}{d(d-1)}G^{(0)}=g_{\mu\nu}^{(0)}n^{(0)\mu}n^{(0)\nu},\\
&\frac{2L^2}{d(d-1)}G^{(2)}=g_{\mu\nu}^{(2)}n^{(0)\mu}n^{(0)\nu}+2g_{\mu\nu}^{(0)}n^{(0)\mu}n^{(1)\nu},\\
&\frac{2L^2}{d(d-1)}G^{(4)}=g_{\mu\nu}^{(4)}n^{(0)\mu}n^{(0)\nu}+2g_{\mu\nu}^{(2)}n^{(0)\mu}n^{(1)\nu}+g_{\mu\nu}^{(0)}n^{(1)\mu}n^{(1)\nu}+2g_{\mu\nu}^{(0)}n^{(0)\mu}n^{(2)\nu}
\end{align}
Comparing $G_{nn}$ in the static and the Vaidya solutions, one gets
\begin{align}\label{lasterm}
\frac{2L^2}{d(d-1)}(G_{nn}^v-G_{nn}^s)=\frac{2L^2}{d(d-1)}(G^{(4)}_v-G^{(4)}_s)=n^{(0)\mu}n^{(0)\nu}(g_{\mu\nu}^{(4),v}-g_{\mu\nu}^{(4),s})+2g_{\mu\nu}^{(0)}n^{(0)\mu}(n^{(2)\nu,v}-n^{(2)\nu,s})
\end{align}
By normality condition of the normal vector, 
\begin{align}
&g_{\mu\nu}n^\mu n^\nu=\epsilon,\\
&g^{(0)}_{\mu\nu}n^{(0)\mu} n^{(0)\nu}=\epsilon,\\
&g^{(2)}_{\mu\nu}n^{(0)\mu} n^{(0)\nu}+2g^{(0)}_{\mu\nu}n^{(0)\mu} n^{(1)\nu}=0,\\
&g^{(4)}_{\mu\nu}n^{(0)\mu} n^{(0)\nu}+2g^{(2)}_{\mu\nu}n^{(0)\mu} n^{(1)\nu}+g^{(0)}_{\mu\nu}n^{(1)\mu} n^{(1)\nu}+2g^{(0)}_{\mu\nu}n^{(0)\mu} n^{(2)\nu}=0.
\end{align}
Hence,
\begin{align}\label{n2}
g^{(0)}_{\mu\nu}n^{(0)\mu} n^{(2)\nu}|_v-g^{(0)}_{\mu\nu}n^{(0)\mu} n^{(2)\nu}|_s=\frac12\left(n^{(0)\mu} n^{(0)\nu}\right)(g^{(4)}_{\mu\nu,v}-g^{(4)}_{\mu\nu,s})
\end{align}
Finally, \eqref{lasterm} by use of \eqref{n2} would be
\begin{align}
\frac{2L^2}{d(d-1)}(G_{nn}^v-G_{nn}^s)=2n^{(0)\mu}n^{(0)\nu}(g_{\mu\nu}^{(4),v}-g_{\mu\nu}^{(4),s})
\end{align}
Setting $g^{(4)}\sim <T>:=T$,
\begin{align}
G^{(4)}_v-G_{s}^{(4)}=\frac{d(d-1)}{L^2}n^{(0)\mu}n^{(0)\nu}(T_{\mu\nu,v}-T_{\mu\nu,s})
\end{align}
In $d+1=5$, $g^{(2)}_{ij}=-P_{ij}$, the Schouten tensor \cite{renor}
\begin{align}
P_{ij}=\frac 12\Big(\bar R_{ij}-\frac{\bar R g^{(0)}_{ij}}{6}\Big)
\end{align}
%----------------------------
\subsection{The FG expansion of the complexity functional}
The complexity functional \eqref{maxv} in FG coordinates is written as
\begin{align}
&\mathcal C_{Any}=\int d^{d-1}\sigma d\rho \frac{1}{2\rho^{\frac{d+1}{2}}}\Big(1+\frac{\rho G_{nn}}{2(d-2)}+\frac{(\rho-\rho_c)K^2}{2(d-1)}+...\Big)\nonumber\\
&.\Big(1+\gamma L^4(\rho^2 \mathscr{C}^2+\rho^2Log\rho \mathcal{A}+\rho^4\mathcal{B}+\rho^6\Big((\nabla_ig^{(4)}_{jk})^2+g^{(4)}_{ij}g^{(4)ij}\Big))\Big),\nonumber\\
&=\int d^{d-1}\sigma d\rho \frac{1}{2\rho^{\frac{d+1}{2}}}\Big(1+\rho C_1+\rho^2C_2+\rho^3C_3+\rho^4C_4+\rho^5C_5+\mathcal O(\rho^6)\Big),
\end{align}
where 
\begin{align}
&C_1=\Big( \frac{ G^{AdS}_{nn}}{2(d-2)}+\frac{K_0^2}{2(d-1)}\Big),\\
&C_2=\gamma L^4\Big( \mathscr{C}^2+Log\rho \mathcal{A}\Big),\\
&C_3=\gamma L^4( \mathscr{C}^2+Log\rho \mathcal{A})(\frac{ G^{AdS}_{nn}}{2(d-2)}+\frac{K_0^2}{2(d-1)})+\Big(\frac{G^{(2)}}{2(d-2)}+\frac{K^2_{(2)}}{2(d-1)}\Big)\\
&C_4=\mathcal B\\
&C_5=\mathcal B\Big( \frac{ G^{AdS}_{nn}}{2(d-2)}+\frac{K_0^2}{2(d-1)}\Big)+\gamma L^4\Big( \mathscr{C}^2+Log\rho \mathcal{A}\Big)\Big(\frac{G^{(2)}}{2(d-2)}+\frac{K^2_{(2)}}{2(d-1)}\Big)\nonumber\\
&+\Big(\frac{G^{(4)}}{2(d-2)}+\frac{K^2_{(4)}}{2(d-1)}\Big)
\end{align}
%----------------------------------------------------------------------

%---------------------------------------------------------+++-----------
\end{document}